# Boundary Conditions of Subharmonic Oscillations in Fixed-Switching-Frequency DC-DC Converters

Chung-Chieh Fang *

## Abstract

Design-oriented boundary conditions for subharmonic oscillations are of great interest recently. Based on a subharmonic oscillation boundary condition reported in a PhD thesis more than a decade ago, extended new boundary conditions are derived in closed forms for general switching DC-DC converters. Sampled-data and harmonic balance analyses are applied and generate equivalent results. It is shown that the equivalent series resistance causes the boundary conditions for voltage/current mode control to have similar forms. Some *recently* reported boundary conditions become *special cases* in view of the general boundary conditions derived. New Nyquist-like design-oriented plots are proposed to predict or prevent the occurrence of the subharmonic oscillation. The relation between the crossover frequency and the subharmonic oscillation is also analyzed.

**KEY WORDS:** Power electronics, DC-DC power conversion, voltage mode control, current mode control, subharmonic oscillation, sampled-data analysis, harmonic balance analysis

---

*C.-C Fang is with Advanced Analog Technology, 2F, No. 17, Industry E. 2nd Rd., Hsinchu 300, Taiwan, Tel: +886-3-5633125 ext 3612, Email: fangcc3@yahoo.com

# Contents

# 1 Introduction

Most power electronics textbooks [1, for example] devote two separate chapters on voltage mode control (VMC) and current mode control (CMC). For CMC, a separate slope-based design is used to avoid subharmonic oscillation. In switching DC-DC converters, three typical local instabilities [2] are subharmonic oscillation (period-doubling bifurcation, fast-scale instability), saddle-node bifurcation [3] (jump or multiple-solution instability), and Neimark-Sacker bifurcation (slow-scale instability). Generally, most averaged models can predict the saddle-node bifurcation and the Neimark-Sacker bifurcation [3], but not the subharmonic oscillation unless the sampling effects are considered as in [4, 5, 6].

The subharmonic oscillation is known to occur when a sampled-data pole crosses -1 in the complex plane [2]. Some *numerical* boundary conditions of subharmonic oscillation based on *simulations* have been reported [7, for example]. Note that, here, the boundary condition means the critical condition in the converter *parameter space*, not about the well known critical eigenvalue at -1 in the *complex plane*. The boundary conditions define the subharmonic oscillation boundaries in the parameter space to separate stable and unstable regions. The numerical boundary conditions have limited usages for the converter design. Boundary conditions in *closed-forms* would greatly facilitate the converter design, because the *quantitative* effect of each *relevant* converter parameter can be clearly seen. Therefore, design-oriented boundary conditions for subharmonic oscillation are of great interest recently. For example, in [1, 8], the boundary condition for CMC with the voltage loop *open* is derived. In [9], a boundary condition in terms of ripple amplitude for a buck converter with a proportional-integral (PI) controller is derived. The PI controller used has only one pole and one zero. These boundary conditions do not consider the effect of equivalent series resistance (ESR).

In [10], a *unified* VMC/CMC block diagram model is proposed, and a *general* closed-form boundary condition for the subharmonic oscillation is derived based on sampled-data analysis. Here, the term "unified" is used because the model is applicable to *both* VMC and CMC, and the term "general" is used because the boundary condition is applicable to all *types* of switching DC-DC converters of any system *dimension* under various *control* schemes. Although the general boundary condition was published in a PhD thesis [10] more than a decade ago, it was not published elsewhere and was not applied to various converters. It is reported here because some of recent results [9, 11] associated with the subharmonic oscillation can be explained in terms of the general boundary condition.

In this paper, based on the general boundary condition, extended design-oriented boundary conditions are derived. The boundary conditions in [1, 8] for CMC with the voltage loop open and the boundary conditions in [9] for a buck converter with a PI controller become *special cases* in view of the general boundary condition. Additionally, the general boundary condition is extended for various converters under various control schemes. For example, it is extended to CMC with the voltage loop *closed*, average current mode control (ACMC) with a type II controller, and VMC with a type III controller. The type II or type III controller, popular in power electronics industry, has an *integrator* to improve the steady-state regulation and has more poles and zeros than a PI controller. The derived boundary conditions show the effects of various converter parameters, such as ESR, loading resistance, feedback gain, and pole/zero locations. It will be shown that the ESR causes the boundary conditions for VMC and CMC to have similar forms. For example, it is well known that the boundary condition for CMC has a term $D - 1/2$, where $D$ is the duty cycle. It is also known [9] that the boundary condition for VMC has a term $1 - 2D + 2D^2$. It will be shown that, with ESR, the boundary condition for VMC or CMC has both terms of $D - 1/2$ and $1 - 2D + 2D^2$. The derived boundary conditions in this paper are expressed in terms of signal *slopes*, instead of signal *amplitude* as in [9]. The slope-based boundary conditions agree coherently with the traditional slope-based analysis for CMC. The boundary conditions can be also expressed in terms of other parameters, such as signal *instantaneous* slope, the source voltage or the compensating ramp slope, for example. It has been reported in [12] that the subharmonic oscillation is unrelated to the ripple *amplitude* for a converter under ACMC. It has also been reported in [13] that a converter can be stabilized by a compensating ramp with increased *instantaneous slope* but with similar *amplitude*.

The analysis of subharmonic oscillation becomes simpler because of the derived closed-form boundary conditions. For a converter under a particular control scheme, once the converter is expressed in terms of the unified VMC/CMC block diagram model, the boundary condition for that particular control scheme can be readily obtained. For example, $V^2$ control [11] is of great interest recently because of its fast response. It will be shown that the boundary condition for $V^2$ control can be easily obtained under the general modeling approach proposed in this paper.

Although sampled-data analysis [14, 15, 16, 17] has been known for many decades, the approach used here is different. First, traditional sampled-data analysis relies heavily on *numerical* analysis to obtain the steady-state solutions and the associated pole stabilities, which gives little insights for the converter *design*. In this paper, the emphasis is on the *closed-form* boundary conditions in the *parameter space*. The effect of each converter parameter can be clearly seen. Second, traditional sampled-data analysis generally starts from an *approximate* sampled-data model, where the accuracy would be lost at the beginning. In this paper, an *exact* sampled-data model is used to derive the boundary conditions which preserve the accuracy and are gradually simplified into various forms.

Harmonic balance modeling [10, 18, 19, 20] is a *complimentary* approach to analyze subharmonic oscillations in switching converters [10, 20]. The modeling approach proposed in [10, 20] is further extended to analyze various control schemes and to derive similar boundary conditions as in the slope-based sampled-data analysis. This paper is originally planned in two parts (sampled-data slope-based analysis and harmonic balance analysis) because of its lengths. However, these two modeling approaches complement and corroborate each other. Both are presented here in a single paper for better cross reference. Note that this paper focuses on the sampled-data slope-based analysis. For some instances, harmonic balance analysis provides additional perspectives on the effects of some particular converter parameters. In this paper, the harmonic balance analysis is applied only to the buck converter. Similar analysis can be applied to other converters. Also note that this paper focuses only on the continuous conduction mode (CCM).

The contributions along with the conclusion are stated at the end of the paper. The remainder of the paper is organized as follows. In Section 2, the operation of VMC and CMC is modeled in a single unified block diagram. In Section 3, steady-state analysis and small-signal analysis are presented. In Section 4, general subharmonic oscillation boundary conditions for buck and boost converters are derived. In Sections 5 and 6, the sampled-data slope-based analysis and the harmonic balance analysis are applied respectively to analyze various control schemes. In Sections 7 and 8, prediction of subharmonic oscillation based on the loop gain and crossover frequency are analyzed.

## 2 Brief Review of Voltage/Current Mode Control Operation

The operation of a general switching DC-DC converter in CCM under *either* VMC or CMC can be described exactly by a unified block diagram model [10, 21, 22] shown in Fig. 1. A unique aspect about Fig. 1 is that the ramps in VMC and CMC are represented by the same $h(t)$. The operation of VMC and CMC is briefly reviewed here to make this paper self-contained. The control (reference) signal $v_r$ controls the output voltage $v_o$ in VMC, and controls the peak inductor current $i_L$ in CMC. Denote the source voltage as $v_s$. In the model, $A_1, A_2 \in \mathbf{R}^{N \times N}$, $B_1, B_2 \in \mathbf{R}^{N \times 2}$, $C, E_1, E_2 \in \mathbf{R}^{1 \times N}$, and $D \in \mathbf{R}^{1 \times 2}$ are constant matrices, where $N$ is the system dimension. For example, $N = 5$ for a buck converter with a type III compensator. Within a clock period $T$, the dynamics is switched between two stages, $S_1$ and $S_2$. Switching occurs when the ramp signal $h(t)$ intersects with the compensator output $y := Cx + Du \in \mathbf{R}$. Denote the ramp amplitude as $V_h$, and denote the switching frequency as $f_s := 1/T$ and let $\omega_s := 2\pi f_s$.

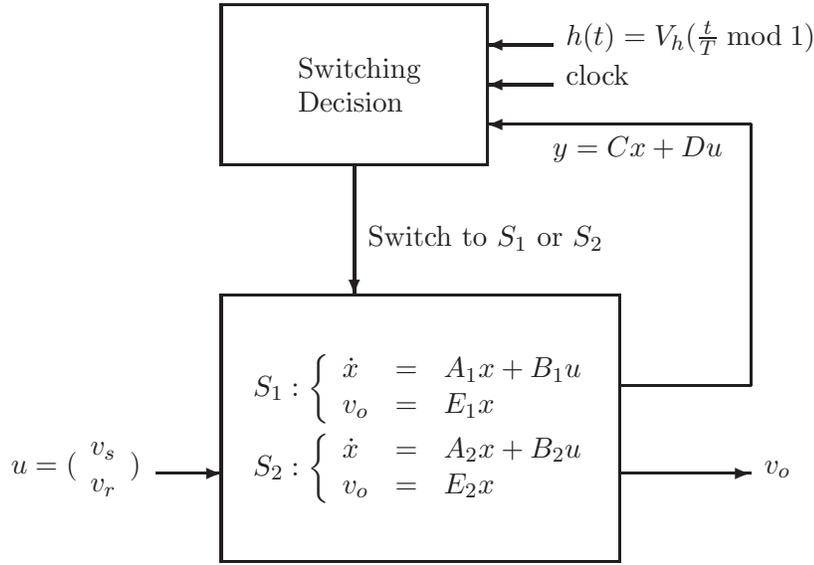

Figure 1: Unified VMC/CMC block diagram model for a switching converter in CCM

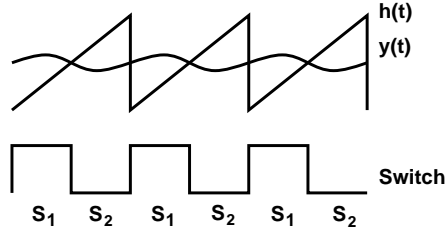

Figure 2: Waveforms for voltage mode control

Typical signal waveforms for VMC and CMC are shown in Fig. 2 and Fig. 3, respectively. In Fig. 3, the ramp has a positive slope, instead of a negative slope as commonly seen in most literatures, in order to be consistent with VMC. Other control schemes (average current mode control, for example) also fit the model of Fig. 1.

## 3 Brief Review of Steady-State and Small-Signal Analysis

Steady-state analysis and small-signal analysis are briefly reviewed here to make this paper self-contained. The periodic solution $x^0(t)$ of the system in Fig. 1 corresponds to a fixed point $x^0(0)$ in the sampled-data dynamics. A typical periodic solution $x^0(t)$ is shown in Fig. 4, where $\dot{x}^0(d^-) = A_1 x^0(d) + B_1 u$ and $\dot{x}^0(d^+) = A_2 x^0(d) + B_2 u$ denote the time derivative of $x^0(t)$ at $t = d^-$ and $d^+$, respectively. Let $y^0(t) = Cx^0(t) + Du$. In steady state, $\dot{y}^0(t) = C\dot{x}^0(t)$. Let the steady-state duty cycle be $D$ and $d := DT$. Confusion of notations for capacitance $C$ and duty cycle $D$ with the matrices $C$ and $D$ can be avoided from the context.

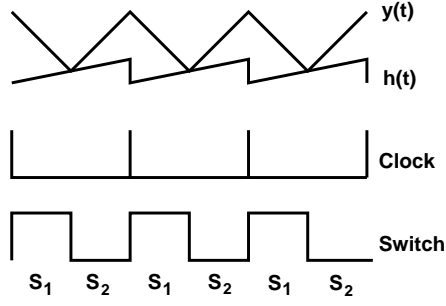

Figure 3: Waveforms for current mode control

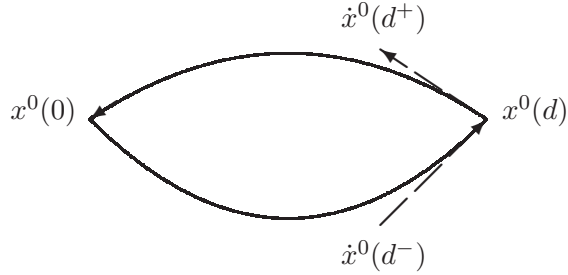

Figure 4: A typical periodic solution $x^0(t)$ of a DC-DC converter in state space

In steady state,

$$x^0(d) = e^{A_1 d} x^0(0) + \int_0^d e^{A_1 \sigma} d\sigma B_1 u \qquad (1)$$

$$x^0(0) = e^{A_2(T-d)} x^0(d) + \int_0^{T-d} e^{A_2 \sigma} d\sigma B_2 u \qquad (2)$$

From (1) and (2), one has

$$x^0(d) = (I - e^{A_1 d} e^{A_2(T-d)})^{-1} (e^{A_1 d} \int_0^{T-d} e^{A_2 \sigma} d\sigma B_2 u + \int_0^d e^{A_1 \sigma} d\sigma B_1 u) \qquad (3)$$

Let $B_1 := [B_{11}, B_{12}]$, $B_2 := [B_{21}, B_{22}]$ to expand the matrices into two columns. The buck converter generally has $A_1 = A_2$, $B_{21} = 0_{N \times 1}$, and $B_{12} = B_{22}$. For $A_1$ and $I - e^{A_1 T}$ being invertible, (3) becomes

$$x^0(d) = (I - e^{A_1 T})^{-1} A_1^{-1} (e^{A_1 d} - I) B_{11} v_s - A_1^{-1} B_{12} v_r \qquad (4)$$

Generally the controller may include an integrator (with a pole at zero), making $A_1$ and $I - e^{A_1 T}$ non-invertible. In that case, the pole at zero can be replaced by a very small number $\delta$, then $A_1$ and $I - e^{A_1 T}$ are invertible. Therefore, the invertibility of $A_1$ or $I - e^{A_1 T}$ is not critical and can be resolved. This statement about invertibility of a matrix is not repeated later.

At the switching instant $t = d$, one has $y^0(d) = Cx^0(d) + Du = h(d)$. Then, from (4),

$$v_s = \frac{h(d) + CA_1^{-1}B_{12}v_r - Du}{C(I - e^{A_1T})^{-1}A_1^{-1}(e^{A_1d} - I)B_{11}} \tag{5}$$

The boost converter generally has $B_1 = B_2$, then

$$x^0(d) = (I - e^{A_1d}e^{A_2(T-d)})^{-1}(e^{A_1d}\int_0^{T-d} e^{A_2\sigma}d\sigma + \int_0^d e^{A_1\sigma}d\sigma)B_1u := X(d)B_1u \tag{6}$$

Using a hat ˆ to denote small perturbations (e.g., $\hat{x}_n = x_n - x^0(0)$), where $x_n$ is the sampled state at $t = nT$. From [10, 21, 23], the linearized sampled-data dynamics is

$$\hat{x}_{n+1} = \Phi\hat{x}_n \tag{7}$$

where

$$\Phi = e^{A_2(T-d)}(I - \frac{(\dot{x}^0(d^-) - \dot{x}^0(d^+))C}{\dot{y}^0(d^-) - \dot{h}(d)})e^{A_1d} \tag{8}$$

Although the *general methodology* of sampled-data analysis has been known in the last three decades, the *closed form* expression of (8) was first published in [10, 23], to the author's knowledge. The closed form of (8) greatly facilitates the derivation of the boundary conditions discussed next.

## 4 Subharmonic Oscillation Boundary Conditions

### 4.1 General Boundary Conditions

The subharmonic oscillation occurs when one eigenvalue of $\Phi$ is $-1$, and $\det[I+\Phi] = 0$. A necessary and sufficient boundary condition for occurrence of subharmonic oscillation is obtained [10, p. 46],

$$\dot{y}^0(d^-) - C(e^{-A_2(T-d)}e^{-A_1d} + I)^{-1}(\dot{x}^0(d^-) - \dot{x}^0(d^+)) = \dot{h}(d) \tag{9}$$

The proof is as follows. Suppose $-1$ is not an eigenvalue of $e^{A_2(T-d)}e^{A_1d}$, then

$$\det[I - \Phi] = \det[I + e^{A_2(T-d)}e^{A_1d}]\det[I - (I + e^{A_2(T-d)}e^{A_1d})^{-1}e^{A_2(T-d)}\frac{\dot{x}^0(d^-) - \dot{x}^0(d^+)}{C\dot{x}^0(d^-) - \dot{h}(d)}Ce^{A_1d}]$$

$$= \det[I + e^{A_2(T-d)}e^{A_1d}][1 - Ce^{A_1d}(I + e^{A_2(T-d)}e^{A_1d})^{-1}e^{A_2(T-d)}\frac{\dot{x}^0(d^-) - \dot{x}^0(d^+)}{\dot{y}^0(d^-) - \dot{h}(d)}]$$

$\det[I + \Phi] = 0$ requires that the last term (inside the second square brackets) of the last equation equals to zero, which leads to (9).

One can expand (9) in terms of $x^0(d)$,

$$C(A_1x^0(d) + B_1u) - C(e^{-A_2(T-d)}e^{-A_1d} + I)^{-1}((A_1 - A_2)x^0(d) + (B_1 - B_2)u) = \dot{h}(d) \tag{10}$$

Note that the condition (9) is valid for *both* VMC and CMC in this unified modeling approach, and it is applicable to *general* switching converters of any system dimension. Also note that in (9), the left side is related to the ripple slopes ($\dot{y}^0(d^-)$, $\dot{x}^0(d^-)$, and $\dot{x}^0(d^+)$), and the right side is the ramp slope $\dot{h}(d)$. As in the popular slope-based boundary condition for the subharmonic oscillation in CMC, the condition (9) is also slope-based.

The condition (9) can be proved to be equivalent to

$$\dot{y}^0(d^+) + C(e^{A_2(T-d)}e^{A_1d} + I)^{-1}(\dot{x}^0(d^-) - \dot{x}^0(d^+)) = \dot{h}(d) \tag{11}$$

Using which one of (9) or (11) depends on convenience. Since the proof for (9) is given, the proof for (11) is omitted to save space.

## 4.2 Buck Converter

The buck converter generally has $A_1 = A_2$, $B_{21} = 0_{N \times 1}$, and $B_{12} = B_{22}$. Using (4), the boundary condition (10) becomes

$$C[(I - e^{A_1 T})^{-1}(e^{A_1 d} - I) + (I + e^{A_1 T})^{-1}]B_{11} = \frac{\dot{h}(d)}{v_s} \tag{12}$$

or in terms of $v_s$, which shows the *critical* value of $v_s$,

$$V_s^* = \frac{\dot{h}(d)}{C[(I - e^{A_1 T})^{-1}(e^{A_1 d} - I) + (I + e^{A_1 T})^{-1}]B_{11}} \tag{13}$$

Subharmonic oscillation is avoided if $v_s < V_s^*$ (if the denominator of (13) is positive as discussed later). The condition (13) is for the trailing edge modulation [24]. For the leading edge modulation, a similar condition has been reported in [10, p. 72].

As will be shown later for the VMC buck converter, the curve (13) as a function of $d = DT$ generally has a minimum at $d = T$ (or equivalently $D = 1$) with a value

$$V_s^*|_{min} = V_s^*|_{d=T} = \frac{\dot{h}(d)}{C[(I + e^{A_1 T})^{-1} - I]B_{11}} \tag{14}$$

Then, for $v_s < V_s^*|_{min}$, the subharmonic oscillation is avoided *for all duty cycles*.

Based on the assumption that the switching frequency $f_s = 1/T$ is much larger than the absolute value of any eigenvalue of $A_1$, matrix approximations such as $e^{A_1 T} \approx I + A_1 T + A_1^2 T^2 / 2$ and $(I + A_1 T)^{-1} \approx I - A_1 T$ can be applied. Then, the boundary condition (12) leads to

$$(\frac{1}{2} - D)CB_{11} - (\frac{1 - 2D + 2D^2}{4})CA_1 B_{11} T \approx \frac{\dot{h}(d)}{v_s} \tag{15}$$

or in terms of $V_s^*$,

$$V_s^* \approx \frac{\dot{h}(d)}{(\frac{1}{2} - D)CB_{11} - (\frac{1 - 2D + 2D^2}{4})CA_1 B_{11} T} \tag{16}$$

Throughout the paper, all approximate boundary conditions are expressed with the approximation sign "$\approx$," and all exact boundary conditions are expressed with the equality sign "$=$."

*Remarks:*

(a) The left side of (15) is a weighted combination of $CB_{11}$ and $CA_1 B_{11} T$. It will be shown that if the equivalent series resistance (ESR) $R_c = 0$, $CA_1 B_{11}$ dominates in VMC, whereas $CB_{11}$ dominates in CMC. For $R_c > 0$, either VMC or CMC has both the terms $CB_{11}$ and $CA_1 B_{11} T$, indicating that the subharmonic oscillation conditions for VMC and CMC are closely related. This result has not been reported.

(b) The boundary condition (15) seems to have a pattern. It is a hypothesis that the *exact* boundary condition has the form

$$C(\sum_{n=0}^{\infty} \delta_n(D) A_1^n T^n) B_{11} = \frac{\dot{h}(d)}{v_s} \tag{17}$$

where $\delta_0(D) = (1 - 2D)/2$, $\delta_1(D) = (-1 + 2D - 2D^2)/4$, etc., and $d\delta_{n+1}(D)/dD = \delta_n(D)$. Since an exact condition as in (12) has been obtained, another exact condition in series expression may give additional insights but may be unnecessary. As a side note, $\delta_2(D) = (-D + 3D^2 - 2D^3)/12$ and $\delta_3(D) = (1 - 2D^2 + 4D^3 - 2D^4)/48$. Note that $d\delta_2(D)/dD \neq \delta_1(D)$, but $d\delta_3(D)/dD = \delta_2(D)$. Further research on the series expression of (17) is pursued and will be reported separately.

(c) The condition (12) is an exact condition, whereas (15) is an approximate one. Based on simulations, if the real controller poles are smaller than $\omega_s/10$, the approximate condition (15) is close to the exact condition (12), and (15) is generally adequate to predict the subharmonic oscillation. If the real controller poles are greater than $\omega_s/10$, using the exact condition (12) gives more accurate results.

(d) The boundary condition (15) does not require a matrix inverse.

## 4.3 Boost Converter

The analysis for the boost converter is similar to that for the buck converter. Let $\Lambda(d) := I + (A_1 - (I + e^{-A_2(T-d)}e^{-A_1 d})^{-1}(A_1 - A_2))X(d)$ to simplify the equation. Using (6), the boundary condition (10) becomes

$$C\Lambda(d)B_1 u = \dot{h}(d) \tag{18}$$

or in terms of $v_s$,

$$V_s^* = \frac{\dot{h}(d) - C\Lambda(d)B_{12}v_r}{C\Lambda(d)B_{11}} \tag{19}$$

Based on (19), it can be proved that, in a boost converter in CCM with proportional feedback VMC (thus the system dimension $N = 2$) and with practical converter parameters, the subharmonic oscillation does not occur. With different modes or control schemes (with a higher system dimension), subharmonic oscillations may still occur in a boost converter. For example, subharmonic oscillation occurs in a boost converter in discontinuous conduction mode [25].

## 4.4 The "S plot": a Slope-Based Plot in the Real Domain

Define an "S plot" as a function of $D = d/T$,

$$S(D) := \dot{y}^0(d^-) - C(e^{-A_2(T-d)}e^{-A_1 d} + I)^{-1}(\dot{x}^0(d^-) - \dot{x}^0(d^+)) \tag{20}$$

$$= \begin{cases} C[(I - e^{A_1 T})^{-1}(e^{A_1 d} - I) + (I + e^{A_1 T})^{-1}]B_{11}v_s & \text{(for the buck converter, from (12))} \\ C\Lambda(DT)B_1 u & \text{(for the boost converter, from (18))} \end{cases} \tag{21}$$

Then, from (9), subharmonic oscillation occurs when

$$S(D) = \dot{h}(d) \tag{22}$$

The S plot facilitates the converter design to avoid the subharmonic oscillation. For example, if the ramp slope $\dot{h}(d)$ is large enough such that $\dot{h}(d) > S(D)$, then the subharmonic oscillation is avoided.

Note that, for the buck converter, $S(T) = C[(I + e^{A_1 T})^{-1} - I]e^{A_1 d}B_{11}v_s \approx -(CB_{11}/2 + CA_1 B_{11}T/4)v_s$. It will be shown later that, for the VMC buck converter, the maximum of $S(D)$ is generally $S(T)$. To avoid the subharmonic oscillation for *all* duty cycle, a ramp slope $\dot{h}(d) > \max S(D)$ is required. It may be over-compensating because such a ramp is applied to avoid the subharmonic oscillation for *all* duty cycle. To avoid the subharmonic oscillation for a particular duty cycle, only $\dot{h}(d) > S(D)$ is required.

### 4.5 Other Approximate Slope-Based Boundary Conditions

In the analysis above, a (first) approximate slope-based boundary condition (15) is presented. Other approximation approaches may be applied to derive similar *slope-based* boundary conditions. Generally these boundary conditions are close to the first approximate boundary condition (15).

#### 4.5.1 Second Approximate Boundary Condition

With a large switching frequency, $(I + e^{A_2(T-d)}e^{A_1 d})^{-1} \approx I/2 - (A_1 d + A_2(T-d))/4$. Then the boundary condition (11) becomes

$$\frac{1}{2}(\dot{y}^0(d^-) + \dot{y}^0(d^+)) - \frac{C}{4}(A_1 d + A_2(T-d))(\dot{x}^0(d^-) - \dot{x}^0(d^+)) \approx \dot{h}(d) \tag{23}$$

This condition is also expressed in terms of signal slopes. Then, a similar condition as in (15) can be obtained.

#### 4.5.2 Third Approximate Boundary Condition under a Large Switching Frequency

Although the switching frequency never reaches infinity, the results obtained under this condition link very well with the well known stability condition for CMC. Note that this section is presented here only to *illustrate a special case* for the boundary condition (9). When the switching frequency is high, $e^{A_1 d} \approx e^{A_2(T-d)} \approx I$. Then the boundary condition (9) becomes

$$\frac{1}{2}(\dot{y}^0(d^-) + \dot{y}^0(d^+)) \approx \dot{h}(d) \tag{24}$$

Like (12), (24) is also a slope-based condition. In CMC, the (inductor current) slopes $\dot{y}^0(d^-)$ and $\dot{y}^0(d^+)$ and the ramp slope $\dot{h}(d)$ are generally expressed in most textbooks [1, p. 448] as $-m_1$, $m_2$, and $m_a$, respectively. Then, (24) corresponds exactly to the well known minimum ramp slope $m_a = (m_2 - m_1)/2$ required to stabilize the converter.

As a side note, the condition (9) or (24) occurs when $\Phi$ has an eigenvalue $-1$. Similar to the proof for (9), one can prove that $\dot{y}^0(d^+) = \dot{h}(d)$ when $\Phi$ has an eigenvalue 0, which generally causes a deadbeat effect. Under CMC, this condition is equivalent to the well known condition $m_2 = m_a$ to have a deadbeat effect [1]. Note that $\dot{y}^0(d^+) = \dot{h}(d)$ is a *general* condition valid for *any* control scheme, not just CMC.

## 5 Application of Sampled-Data Slope-Based Analysis

Without loss of generality, the sampled-data slope-based analysis is applied to buck converters. Analysis of Boost converters can be applied similarly. The exact condition (12) and the approximate condition (15) are applied to analyze various VMC/CMC control schemes presented next. The compensator complexity increases from a simple proportional compensator to a type III compensator. The system dimension $N$ increases from two to five.

### 5.1 Proportional Voltage Mode Control (PVMC)

Without loss of generality, let the voltage loop have a proportional feedback gain $k_p$. One has $y = k_p(v_r - v_o)$, as shown in Fig. 5. For designation purpose, this control scheme is called proportional VMC (PVMC).

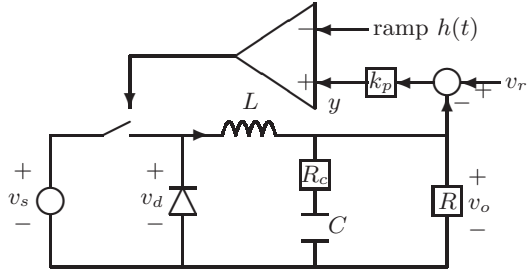

Figure 5: A buck converter under PVMC

In the power stage, let the state be $x = (i_L, v_C)'$, where $i_L$ is the inductor current and $v_C$ is the capacitor voltage. Let the load be $R$, the inductance be $L$, the capacitance be $C$, and the equivalent series resistance (ESR) be $R_c$. Let $\rho = R/(R + R_c)$. For $R_c = 0$, $\rho = 1$. Then,

$$A_1 = A_2 = \rho \begin{bmatrix} \frac{-R_c}{L} & \frac{-1}{L} \\ \frac{1}{C} & \frac{-1}{RC} \end{bmatrix} \tag{25}$$

$$B_1 = [B_{11}, B_{12}] = \begin{bmatrix} \frac{1}{L} & 0 \\ 0 & 0 \end{bmatrix}, \quad B_2 = [B_{21}, B_{22}] = \begin{bmatrix} 0 & 0 \\ 0 & 0 \end{bmatrix} \tag{26}$$

$$E_1 = E_2 = \rho \begin{bmatrix} R_c & 1 \end{bmatrix} \tag{27}$$

$$\begin{array}{ll}
C = -k_p \rho [R_c, 1] & D = [0, k_p] \\
CB_{11} = \frac{-k_p \rho R_c}{L} & CA_1 B_{11} = \frac{-k_p \rho^2}{LC}(1 - \frac{R_c^2 C}{L}) \approx \frac{-k_p}{LC} \\
\text{For } R_c = 0, CB_{11} = 0 & CA_1 B_{11} = \frac{-k_p}{LC}
\end{array}$$

Then, from (16),

$$V_s^* \approx \frac{4V_h LC}{\rho k_p T^2} \left( \frac{1}{\frac{4R_c C}{T}(D - \frac{1}{2}) + \rho(1 - \frac{R_c^2 C}{L})(1 - 2D + 2D^2)} \right) \tag{28}$$

$$\approx \frac{4V_h LC}{\rho k_p T^2} \left( \frac{1}{\frac{4R_c C}{T}(D - \frac{1}{2}) + (1 - 2D + 2D^2)} \right) \tag{29}$$

For $R_c = 0$,

$$V_s^* \approx \frac{4V_h LC}{k_p T^2} \left( \frac{1}{1 - 2D + 2D^2} \right) \tag{30}$$

In [9], a similar condition (only for $R_c = 0$) with a minor correction term is obtained. Compared with (30), (28) has an additional term involving $D - 1/2$, which is related to CMC when the subharmonic oscillation occurs as discussed next. For $R_c C \ll T$, (30) is a good estimate of the critical source voltage. If the condition $R_c C \ll T$ is not met, (28) will be more accurate than (30).

Both (28) and (30) are functions of $D$, and $D$ is chosen to meet the required output voltage. For example, $v_o \approx v_s D$ for the open-loop buck converter. Given a value of $D$, (28) or (30) sets the maximum value of $v_s$ to avoid the subharmonic oscillation. In a closed loop converter, $D$ is

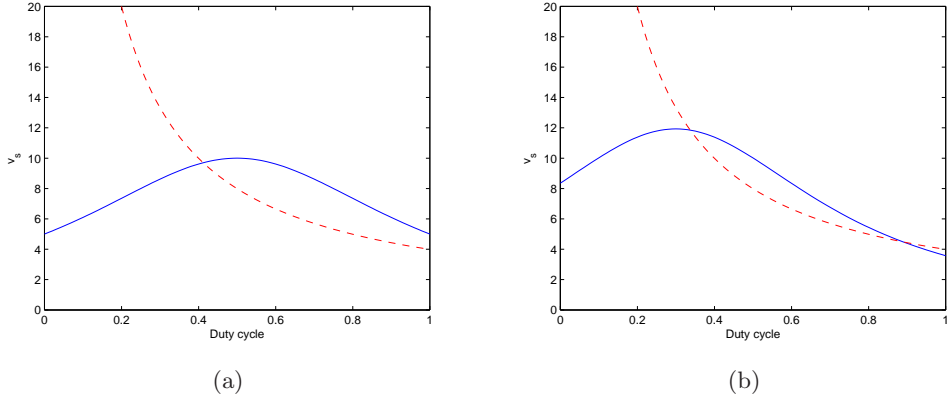

(a)                                            (b)

Figure 6: Plots of (28) (solid line) and (31) (dashed line), and the intersection shows $V_s^*$. (a) $R_c = 0$, (b) $R_c = 2$ mΩ.

determined by (5), or by the steady-state equation, $DV_h = k_p(v_r - v_o) = k_p(v_r - v_s D)$, rearranged as

$$v_s = \frac{v_r}{D} - \frac{V_h}{k_p} \tag{31}$$

For a *practical* converter, $k_p$ is large and $v_s \approx v_r/D$. When the subharmonic oscillation occurs, both the steady-state condition (31) and the subharmonic oscillation condition (28) (or (30)) need to be met. Therefore, the intersection of (28) (or (30)) with (31) is the critical source voltage in a closed loop converter.

Another way without plotting the two curves (28) and (31) to determine the critical value for the subharmonic oscillation is as follows. One way is to subtract (31) by (28). Another way is to divide (31) by (28), which leads to an "M plot" as discussed in Sec.7.

**Example 1.** Consider a PVMC buck converter with $k_p = 80$. The converter parameters are $V_h = 1$, $v_r = 4$, $f_s = 1$ MHz, $L = 1$ μH, $C = 100$ μF, and $R = 2$ Ω.

To see the effects of $R_c$, the curves of (28) and (31) for different values of $R_c$ are also shown in Fig. 6. The intersection of (28) and (31) determines $V_s^*$. The converter is stable for those operating range of $D$ such that the curve (31) is below the curve (28). It can be shown that the curve (28) is actually almost identical to the curve (13), omitted for brevity.

For $R_c = 0$, the curves of (28) and (31) are shown in Fig. 6(a), intersecting at $(D, v_s) = (0.41, 9.7)$. The stable operating range of $D$ is $[0.41, 1]$. Simulation (Fig. 7) with $v_s = 10 > V_s^*$ shows the subharmonic oscillation as predicted.

For $R_c = 2$ mΩ, the curves of (28) and (31) are shown in Fig. 6(b), intersecting at two points $(D, v_s) = (0.34, 11.85)$ and $(0.89, 4.48)$. The curve (28) is shifted to the upper-left (because the term $(4R_c C/T)(D - 1/2)$ in (28) becomes significant), and $V_s^*$ is increased to 11.85. Simulation (Fig. 8) with $v_s = 12.5 > V_s^*$ also shows the subharmonic oscillation as predicted. The stable operating range of $D$ is $[0.34, 0.89]$. For $v_s = 4.44$ (and $D = 0.89$), another subharmonic oscillation occurs, along with a border-collision bifurcation, because $y(t)$ is out of bounds of $h(t)$. The signal plot is omitted for brevity. □

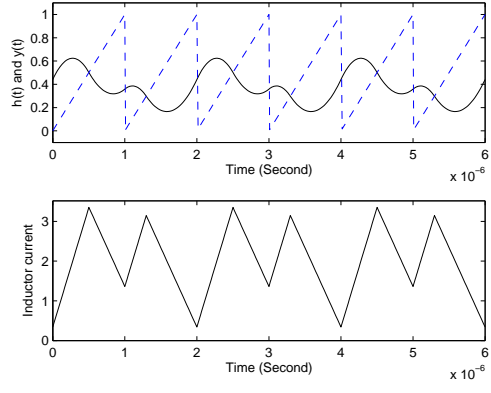

Figure 7: Plots of $y(t)$ and $i_L(t)$ (solid lines), and $h(t)$ (dashed line), $R_c = 0$

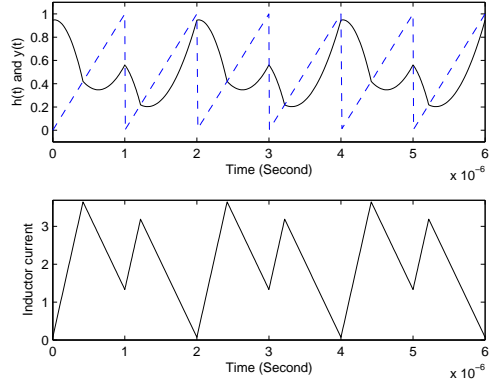

Figure 8: Plots of $y(t)$ and $i_L(t)$ (solid lines), and $h(t)$ (dashed line), $R_c = 2$ mΩ

## 5.2 $V^2$ Control: Similar to PVMC

The constant-frequency peak voltage regulator (CF-PVR, as shown in Fig. 9), a type of $V^2$ control [11], is proved below to be a special case of PVMC in terms of the boundary condition. In CF-PVR, the output voltage is sensed (through a voltage divider with a gain of $k_p$), added with a stabilization ramp $h(t)$, and compared with a reference signal $v_r$ to determine the duty cycle. In terms of Fig. 1, $y = v_r - k_p v_o$, and the matrices $C = -k_p \rho [R_c, 1]$ (same as for PVMC) and $D = [0, 1]$.

Since, in terms of Fig. 1, the model is almost the same as PVMC, the boundary condition (29) for the PVMC buck converter is also applicable to CF-PVR. Assume $R_c \ll R$, then $\rho \approx 1$. From (29), the subharmonic oscillation is avoided if

$$v_s < \frac{4V_h LC}{k_p T^2} \left( \frac{1}{\frac{4R_c C}{T}(D - \frac{1}{2}) + (1 - 2D + 2D^2)} \right) \tag{32}$$

With the fact that $v_o = D v_s$ for the buck converter, one can rearrange (32) as

$$\frac{V_h L}{k_p v_o R_c T} > \frac{2D - 1}{2D} + \frac{T}{R_c C}\left(\frac{1 - 2D}{4D} + \frac{D}{2}\right) \tag{33}$$

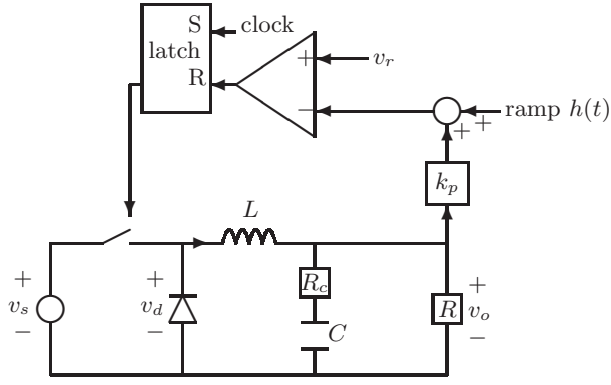

Figure 9: A buck converter under CF-PVR

which agrees with [11, Eq. 7] (by setting $k_p = 1$) and shows the required ramp amplitude $V_h$ to stabilize the subharmonic oscillation. Without the ramp ($V_h = 0$), rearranging (32), subharmonic oscillation is avoided if

$$\frac{T}{R_c C} < \frac{1}{\frac{1}{2} + \frac{D^2}{1-2D}} \tag{34}$$

or, equivalently,

$$\frac{R_c C}{T} > \frac{1}{2} + \frac{D^2}{1-2D} \text{ and } D < \frac{1}{2} \tag{35}$$

also agreed with [11, Eq. 5].

*Remarks:*

(a) The conditions (33) is applicable to *both* PVMC and CF-PVR, and (34) or (35) is a special case ($V_h = 0$) of (33).

(b) $D < 1/2$ is explicitly required in (35), whereas $D < 1/2$ is implicitly required in (34).

(c) The subharmonic oscillation for CF-PVR does not necessarily occur at $D = 1/2$ as claimed in [26]. Based on (35), if $R_c C = 5T/8$, for example, the subharmonic oscillation occur at $D = 1/4$. A *single* pole is shown in [26, p. 352] to be unstable when $D > 1/2$, it cannot be a subharmonic *oscillation* because a pair of unstable complex poles is required.

### 5.3 CMC With the Voltage Loop Open

In CMC with the voltage loop open, as shown in Fig. 10, a switching occurs when $i_c - i_L(t) = h(t)$, where $i_c$ (denoted as $v_r$ in Fig. 1) is a peak inductor current control signal. One has $y = -i_L + i_c$ and

$$\begin{array}{ll} C = [-1, 0] & D = [0, 1] \\ CB_{11} = \frac{-1}{L} & CA_1 B_{11} = \frac{\rho R_c}{L^2} \\ \text{For } R_c = 0, \; CB_{11} = \frac{-1}{L} & CA_1 B_{11} = 0 \end{array}$$

Note that in CMC, $CB_{11}$ dominates (much greater than $CA_1 B_{11} T$ in absolute value). Let the ramp slope be $m_a = \dot{h}(d)$, and let the critical (minimum) ramp slope required to stabilize the converter be $m_a^*$.

**Case I:** $R_c = 0$.

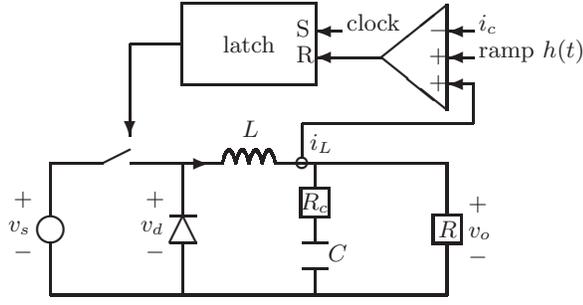

Figure 10: A buck converter under current-mode control with the voltage loop open

For $R_c = 0$, from (15), subharmonic oscillation is avoided [8] if

$$m_a > m_a^* \approx \frac{v_s}{L}(D - \frac{1}{2}) \quad (36)$$

Without the compensating ramp ($m_a = 0$), the converter is unstable for $D > 0.5$ as the condition (36) is violated (because the left side of (36) is zero whereas the right side of (36) is positive). The condition (36) defines the minimum of $m_a$ and $L$, and the maximum of $v_s$ and $D$ to avoid the subharmonic oscillation. For example, the maximum of $v_s$ to avoid the subharmonic oscillation is $m_a L/(D - 1/2)$, which shows that the stable operation range of the source voltage $v_s$ is linearly proportional to the ramp slope. Adding the compensating ramp not only improves the stability but also enlarges the operating range of the source voltage (line regulation).

**Case II:** $R_c > 0$.
For CMC with $R_c > 0$, the boundary condition (15) becomes

$$m_a^* \approx \frac{v_s}{L}(D - \frac{1}{2} - \frac{\rho R_c T}{L}(\frac{1 - 2D + 2D^2}{4})) \quad (37)$$

Compared with (36), (37) has an additional term $1 - 2D + 2D^2$ related to VMC as discussed above. Without the compensating ramp ($m_a = 0$), (37) is a quadratic equation of $D$. Ignoring the second solution greater than one, the critical value of $D$ when the subharmonic oscillation occurs is

$$\begin{aligned} D^* &\approx \frac{1}{2} + \frac{L}{R_c T} - \frac{L}{R_c T}\sqrt{1 - \frac{R_c^2 T^2}{4L^2}} \\ &\approx \frac{1}{2} + \frac{\rho R_c T}{8L} \text{ (for } R_c T \ll L\text{)} \end{aligned} \quad (38)$$

The subharmonic oscillation is generally believed to occur when $D = 1/2$ without the ramp. However, from (38), the ESR extends the operating range of the duty cycle beyond $1/2$. Generally, $\rho R_c T/8L$ is small, and the instability at $D = 1/2$ is still a good criterion.

The boundary conditions can be also rearranged in terms of $v_s$. From (36) and (37), for $R_c = 0$ and $R_c > 0$, respectively,

$$V_s^* \approx \frac{V_h L}{T}(\frac{1}{D - \frac{1}{2}}) \quad (39)$$

$$V_s^* \approx \frac{V_h L}{T}(\frac{1}{D - \frac{1}{2} - \frac{\rho R_c T}{L}(\frac{1-2D+2D^2}{4})}) \quad (40)$$

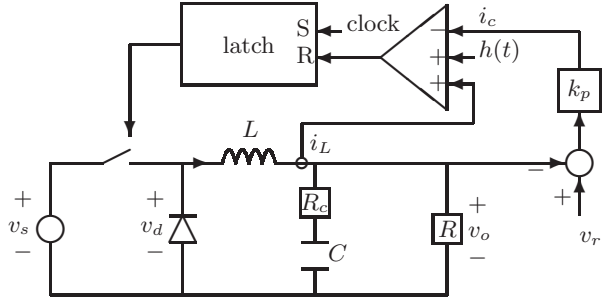

Figure 11: A buck converter under current-mode control with the voltage loop closed

*Remarks:*

(a) Care should be taken about the direction of instability. In an equality, when the both sides are divided by a negative number, the inequality sign is reversed. Assume that the ramp is very small, for $D > 1/2$, the converter is stable if $v_s < V_s^*$. For $D < 1/2$, the converter is stable if $v_s > V_s^*$.

(b) Comparing these two equations with (28) and (30) for VMC, one can see that the boundary conditions for VMC and CMC have very similar forms. Both have the terms $D - 1/2$ and $1 - 2D + 2D^2$ for $R_c > 0$.

## 5.4 CMC With the Voltage Loop Closed

Without loss of generality, let the voltage loop have proportional feedback $k_p(v_r - v_o)$, as shown in Fig. 11. One has $y = k_p(v_r - v_o) - i_L$ and

$$C = -[1 + k_p \rho R_c, k_p \rho] \qquad D = [0, k_p]$$
$$CB_{11} = \tfrac{-1}{L}(k_p \rho R_c + 1) \qquad CA_1 B_{11} = \tfrac{-k_p \rho^2}{LC}(1 - \tfrac{R_c^2 C}{L}) + \tfrac{\rho R_c}{L^2}$$
$$\text{For } R_c = 0,\ CB_{11} = \tfrac{-1}{L} \qquad CA_1 B_{11} = \tfrac{-k_p}{LC}$$

The boundary condition (15) rearranged in terms of the critical feedback gain $k_p^*$ is

$$k_p^* \approx \frac{\frac{m_a L}{v_s} + \frac{\rho R_c T}{L}(\frac{1 - 2D + 2D^2}{4}) - D + \frac{1}{2}}{\frac{\rho^2 T}{C}(1 - \frac{R_c^2 C}{L})(\frac{1 - 2D + 2D^2}{4}) + (D - \frac{1}{2})\rho R_c} \tag{41}$$

For a special case $R_c = 0$, one has

$$k_p^* \approx \frac{\frac{m_a L}{v_s} - D + \frac{1}{2}}{\frac{T}{C}(\frac{1 - 2D + 2D^2}{4})} \tag{42}$$

agreed with [27, Eq. 30]. A larger $T/C$ leads to a smaller $k_p^*$. Also, as $D$ is closer to $1/2$, $(1 - 2D)/(1 - 2D + 2D^2)$ and hence $k_p^*$ become smaller.

Subharmonic oscillation is avoided if $k_p < k_p^*$. The *gain margin to avoid the subharmonic oscillation* is $20 \log(k_p^*)$. The gain margin based on traditional averaged analysis generally helps to avoid the Neimark-Sacker bifurcation, but not the subharmonic oscillation, as shown in the next example.

The traditional CMC design is as follows. Choose a ramp slope to ensure the current loop is stable. Based on an averaged model, one derives the control-to-output transfer function and designs a voltage loop controller to have enough phase and gain margins. It is possible that even with current loop designed to be stabilized by the compensating ramp, the subharmonic oscillation still occurs due to the voltage loop.

**Example 2.** (*Slope-based analysis can predict the critical value of feedback gain $k_p^*$ to avoid the subharmonic oscillation, whereas the averaged models fail to predict it.*) Consider a CMC buck converter with proportional control (compensator transfer function $G_c(s) = k_p$) in the voltage loop. The converter parameters are $f_s = 300$ kHz, $L = 900$ nH, $C = 990$ $\mu$F, $R = 0.4$ $\Omega$, and $v_r = 3.34$ V. The source voltage is 5.5 V, and the duty cycle $D$ is around 0.6. Since $D > 0.5$, the converter is unstable if a compensating ramp is not added. Adding a compensating ramp $m_a = m_2/2 = v_s D/2L = 1.8333 \times 10^6$ V/s is supposed to stabilize the converter for all duty cycles, where $m_2$ is the inductor current slope during the off stage. One has $V_h = m_a T = 6.11$. However, with the voltage loop closed, the subharmonic oscillation may still occur even though the current loop is supposed to be stabilized by the ramp.

The ESR $R_c$ contributes to the output voltage ripple, and hence affects the current loop. Two cases are simulated: $R_c = 5$ m$\Omega$ and $R_c = 0$. Simulations show that with $R_c = 5$ m$\Omega$, the critical gain $k_p^*$ is 237, and with $R_c = 0$, $k_p^* = 452$, which are also confirmed by the exact sampled-data analysis [21].

Take $R_c = 5$ m$\Omega$ and $k_p = 237$, for example. The signal waveforms indicating (weak) subharmonic oscillation are shown in Fig. 12. The steady states are $2T$-periodic orbits. The *unstable* $T$-periodic orbit $y^0(t) = Cx^0(t) + Du = k_p(v_r - v_o(t)) - i_L(t)$ (with $i_L$ inverted) still exists and is shown in Fig. 13, along with $h(t)$. If the $T$-periodic orbit $x^0(t)$ is perturbed, it will go to the stable $2T$-periodic orbit shown in Fig. 12. The designed slopes of (on-time inductor current slope) $m_1 = (1-D)v_s/L$ and (off-time inductor current slope) $m_2 = Dv_s/L$ are $2.48 \times 10^6$ and $3.63 \times 10^6$, respectively. However, ESR $R_c$ contributes to the output voltage ripple, and the actual slopes of $y = k_p(v_r - v_o) - i_L$ shown in Fig. 13 are $5.39 \times 10^6$ and $7.88 \times 10^6$, respectively. Even with increased ripples, $(\dot{y}^0(d^+) + m_a)/(\dot{y}^0(d^-) + m_a) = -0.84 > -1$, and the current loop should be still stable according to the averaged models. However, the subharmonic oscillation still occurs as shown in Fig. 12.

The critical gain $k_p^*$ can be predicted by (41). From (41), $k_p^* = 223$ for $R_c = 5$ m$\Omega$, and $k_p^* = 468$ for $R_c = 0$, agreed closely with the simulation results and the exact sampled-data analysis. The small discrepancy is due to the error of the duty cycle. With the voltage loop closed, the actual value of the duty cycle from the simulation for $R_c = 5$ m$\Omega$ is 0.5941, not 0.6 as supposed. With the actual value of the duty cycle of 0.5941, from (41), one has $k_p^* = 237$, agreed exactly with the simulation results.

Two averaged models are used for comparison. One is from [4], and the other is from [1, p. 470]. The control-to-output ($i_c$-to-$v_o$) frequency responses are, respectively, also shown in Fig. 14,

$$\frac{4441321980(s + 2.02 \times 10^5)}{(s + 3273)(s^2 + 5.92 \times 10^5 s + 8.883 \times 10^{11})} \tag{43}$$

$$\frac{5000(s + 2.02 \times 10^5)}{(s + 3276)(s + 9.993 \times 10^5)} \tag{44}$$

Both show infinite gain margins, and do not accurately predict the critical gain of 237 to avoid the subharmonic oscillation. Both models have a similar ESR zero and a similar low frequency pole, whereas the model of [4] has a pair of complex poles associated with half the switching frequency as the resonance frequency instead of a single pole as in [1].

With $k_p = 237$, the loop gains ($i_c$-to-$v_o k_p$) of the two averaged models are shown in Fig. 15. The averaged model of [4] shows that the phase margin is 36.5 degrees. However, the subharmonic oscillation still occurs. The crossover frequency is $w_c = 1.28 \times 10^6$ rad/s, and $w_c/\omega_s = 0.68$. It will be shown in Section 8 that a large crossover frequency likely leads to the subharmonic oscillation.

The averaged model of [1] even shows a phase margin of 129 degrees. However, the subharmonic oscillation still occurs. The crossover frequency is $w_c = 7.19 \times 10^5$ rad/s, and $w_c/\omega_s = 0.3814$ is also large as will be discussed in Section 8. □

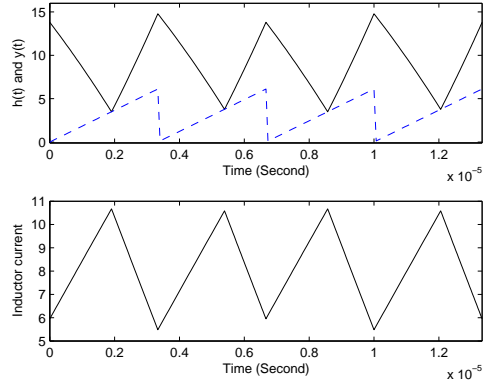

Figure 12: Signal waveforms showing (weak) subharmonic oscillation, $R_c = 5$ mΩ and $k_p = 237$

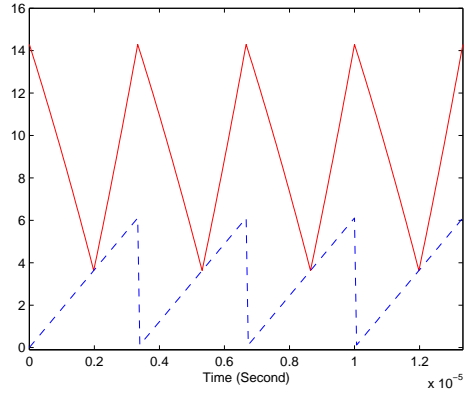

Figure 13: Unstable orbit $y^0(t)$ (solid line) and $h(t)$ (dashed line), $R_c = 5$ mΩ and $k_p = 237$

## 5.5 Enhanced $V^2$ Control: Similar to CMC With the Voltage Loop Closed

Compared with CF-PVR, the enhanced $V^2$ control [26] has an additional loop from the inductor current through a sensing resistor $R_i$ shown in Fig. 16. One has $y = v_r - v_o - R_i i_L$. Similar to

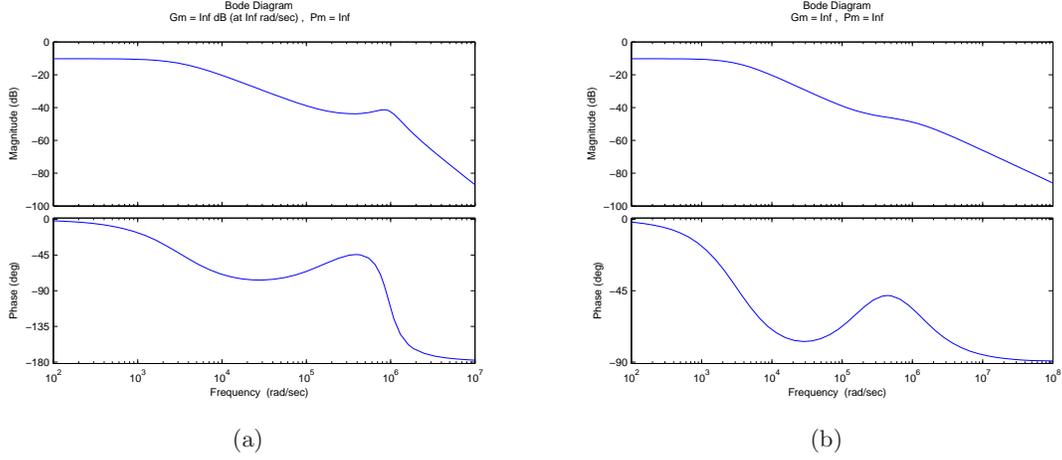

Figure 14: Bode plots for (a) averaged model of [4], (b) averaged model of [1]

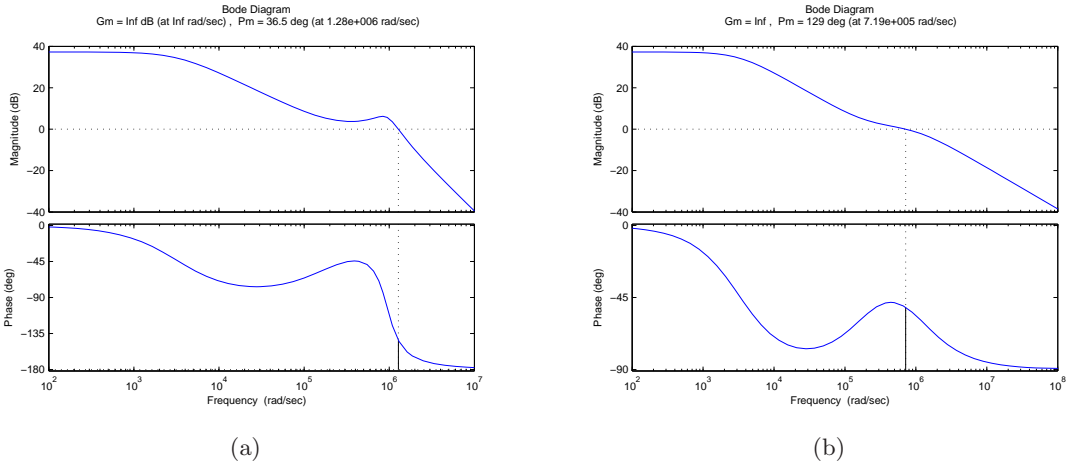

Figure 15: Loop gains for (a) averaged model of [4] and (b) averaged model of [1]

CMC with the voltage loop closed and $k_p = 1$, one has

$$C = -[\rho R_c + R_i, \rho] \qquad D = [0, 1]$$
$$CB_{11} = \frac{-1}{L}(\rho R_c + R_i) \qquad CA_1 B_{11} = \frac{-\rho^2}{LC}(1 - \frac{R_c(\rho R_c + R_i)C}{\rho L})$$

Consider a special set of converter parameters (with large capacitance $C$ and small ESR $R_c$) [26, p. 351], for example. If $L/R_c \gg (R_c + R_i)C \gg T$, then $CA_1 B_{11} \gg CA_1^2 B_{11} T$ and the subharmonic oscillation occurs (without the ramp, $h(t) = 0$) around $D = 1/2$, agreed with [26, p. 351]. Similarly to the $V^2$ control, the subharmonic oscillation does not occur exactly at $D = 1/2$. For other set of converter parameters, the analysis is similar to CMC with the voltage loop closed and is omitted here to save space.

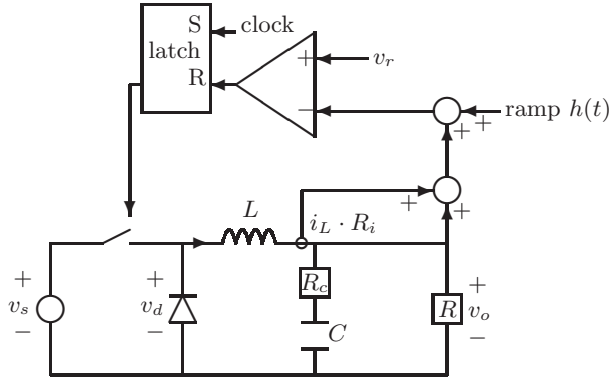

Figure 16: A buck converter under enhanced $V^2$ control

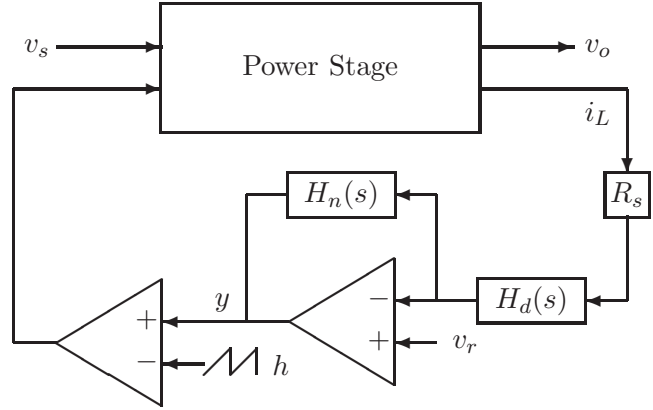

Figure 17: System diagram of an ACMC DC-DC converter

## 5.6 Average Current Mode Control (ACMC) with a Type II Compensator

An ACMC DC-DC converter is shown in Fig. 17. The operation of ACMC is as follows [28]: The inductor current $i_L$ is sensed by a resistor $R_s$ and compared with a voltage reference $v_r$ from a voltage loop (not shown). The difference is amplified by a current-loop compensator, generally a type II compensator [29, p. 256],

$$G_c(s) := \frac{H_n(s)}{H_d(s)} = \frac{K_c(1+\frac{s}{\omega_z})}{(s+\delta)(1+\frac{s}{\omega_p})} \tag{45}$$

The compensator has poles at $-\delta$ and $-\omega_p$, where $\delta$ is a small number close to zero. For $\delta = 0$, the compensator has an integrator with a pole at zero. The compensator output $y(t)$ is compared with the ramp $h(t)$ to determine the duty cycle. Let the state be $x = (i_L, v_C, v_{e1}, v_{e2})'$, where $v_{e1}$ and $v_{e2}$ are the states of the current-loop compensator. Then,

$$A_1 = A_2 = \begin{bmatrix} \frac{-\rho R_c}{L} & \frac{-\rho}{L} & 0 & 0 \\ \frac{\rho}{C} & \frac{-\rho}{RC} & 0 & 0 \\ 0 & 0 & 0 & 1 \\ -\omega_p R_s & 0 & -\delta\omega_p & -\delta - \omega_p \end{bmatrix}$$

$$B_1 = \begin{bmatrix} \frac{1}{L} & 0 \\ 0 & 0 \\ 0 & 0 \\ 0 & \omega_p \end{bmatrix}, \quad B_2 = \begin{bmatrix} 0 & 0 \\ 0 & 0 \\ 0 & 0 \\ 0 & \omega_p \end{bmatrix}$$

$$C = \begin{bmatrix} 0 & 0 & K_c & \frac{K_c}{\omega_z} \end{bmatrix}, \quad D = \begin{bmatrix} 0 & 1 \end{bmatrix}$$

$$E_1 = E_2 = \begin{bmatrix} \rho R_c & \rho & 0 & 0 \end{bmatrix}$$

Since the integrator is modeled as $1/(s+\delta)$, both $A_1$ and $A_2$ are invertible.

**Case I:** $\omega_p > \omega_s/10$.

In this case, the compensator pole $\omega_p$ is close to $\omega_s$, and those higher order terms ($CA_1^2 B_{11} T^2$, for example) in (17) may be larger than $CB_{11}$ and $CA_1 B_{11} T$, and one needs to use the exact condition (13). Generally, the compensator has a pole at zero in the complex plane, making $(I - e^{A_1 T})^{-1}$ singular in (13). It can be easily resolved by using a very small number for the zero pole as discussed previously.

**Case II:** $\omega_p < \omega_s/10$.

Based on simulations as shown in the next example, for $\omega_p < \omega_s/10$, the boundary condition (16) is still a good approximation. Here, $CB_{11} = 0$ and $CA_1 B_{11} = -K_c R_s \omega_p/\omega_z L$, then the boundary condition (16) becomes

$$V_s^* \approx \frac{4V_h \omega_z L}{T^2 K_c R_s \omega_p}\left(\frac{1}{1 - 2D + 2D^2}\right) \tag{46}$$

**Example 3.** Consider a buck converter under ACMC [30, p. 114]. The power stage parameters are $v_s = 14$ V, $v_o = 5$ V, $v_r = 0.5$, $V_h = 1$, $f_s = 50$ kHz, $L = 46.1$ $\mu$H, $C = 380$ $\mu$F with ESR $R_c = 0.02$ $\Omega$, and $R = 1$ $\Omega$. The inductor current sensing resistance is $R_s = 0.1$ $\Omega$. The compensator has a zero at $\omega_z = 5652.9$ rad/s, two poles at 0 and $\omega_p$, and a gain $K_c = 75506$.

The compensator pole $\omega_p$ is varied from $0.14\omega_s$ to $0.81\omega_s$. An unstable *window* of $\omega_p$ between $0.18\omega_s$ and $0.49\omega_s$ was found and reported in [12]. When $\omega_p$ is inside the window, the subharmonic oscillation occurs. Take $\omega_p = 0.49\omega_s$, for example. The converter is unstable. The signal waveform indicating the subharmonic oscillation is shown in Fig. 18. The unstable $T$-periodic orbit is shown in Fig. 19. When the $T$-periodic orbit is perturbed, it will lead to the $2T$-periodic orbit shown in Fig. 18. The unstable window of $\omega_p$ can be predicted exactly by (13), shown in Fig. 20.

In the above analysis, $v_s$ is fixed at 14 V, and $\omega_p$ is varied. Next, let $\omega_p$ be fixed at $\omega_s/10$, and $v_s$ is varied to determine $V_s^*$. Since $\omega_p$ is smaller, the approximate condition (46) is close to the exact condition (13) as shown in Fig. 21, indicating $V_s^* = 19$. From (46), a smaller $\omega_p$ results in a larger $V_s^*$ as expected.

In ACMC, the subharmonic oscillation is unrelated to the ripple *amplitude* [12]. Take $\omega_p = 0.81\omega_s$, for example. The converter is stable. The stable $T$-periodic orbit is shown in Fig. 22. In Fig. 22, the signal $y^0(t)$ has a large signal amplitude but is still stable, whereas in Fig. 19, $y^0(t)$ has a smaller signal amplitude but is unstable. □

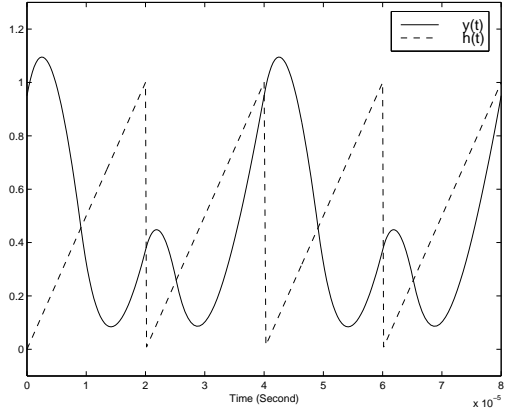

Figure 18: Stable $2T$-periodic orbit, $\omega_p = 0.49\omega_s$

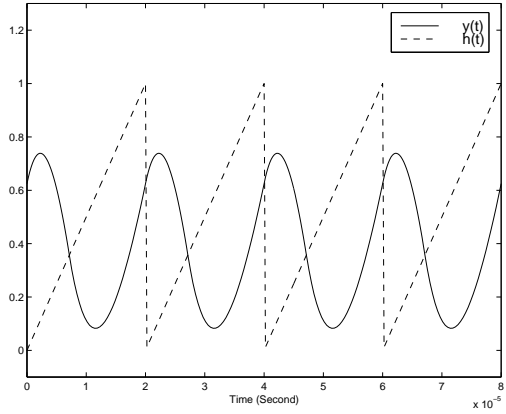

Figure 19: Unstable $T$-periodic $y^0(t)$, $\omega_p = 0.49\omega_s$

## 5.7 Average Current Mode Control (ACMC) with a PI Compensator

Instead of a type II compensator, consider an ACMC buck converter with a PI compensator,

$$G_c(s) := \frac{H_n(s)}{H_d(s)} = \frac{K_c(1 + \frac{s}{\omega_z})}{(s + \delta)} \qquad (47)$$

Then,

$$
\begin{aligned}
A_1 &= A_2 = \begin{bmatrix} \frac{-\rho R_c}{L} & \frac{-\rho}{L} & 0 \\ \frac{\rho}{C} & \frac{-\rho}{RC} & 0 \\ -R_s & 0 & -\delta \end{bmatrix} \\
B_1 &= \begin{bmatrix} \frac{1}{L} & 0 \\ 0 & 0 \\ 0 & 1 \end{bmatrix}, \quad B_2 = \begin{bmatrix} 0 & 0 \\ 0 & 0 \\ 0 & 1 \end{bmatrix} \\
C &= \begin{bmatrix} -\frac{R_s K_c}{\omega_z} & 0 & K_c(1 - \frac{\delta}{\omega_z}) \end{bmatrix}, \quad D = \begin{bmatrix} 0 & 1 + \frac{K_c}{\omega_z} \end{bmatrix} \\
E_1 &= E_2 = \begin{bmatrix} \rho R_c & \rho & 0 \end{bmatrix}
\end{aligned}
$$

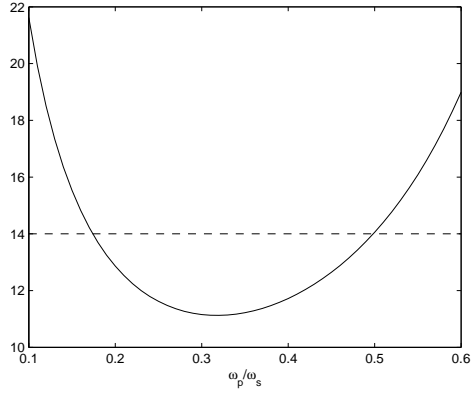

Figure 20: Plot of (13), intersecting with $v_s = 14$ at $\omega_p/\omega_s = 0.18$ and $0.49$, indicating the unstable window of $\omega_p$

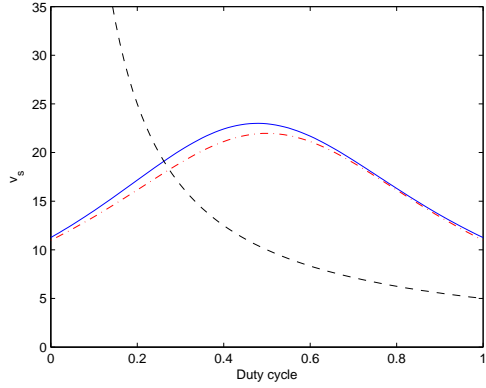

Figure 21: Approximate condition (46) (dash-dotted line) is close to the exact condition (13) (solid line), $V_s^* = 19$

Since the integrator is modeled as $1/(s+\delta)$, both $A_1$ and $A_2$ are invertible. One has (with $\delta \approx 0$)

$$CB_{11} = -\frac{R_s K_c}{\omega_z L}, \qquad CA_1 B_{11} = \frac{R_s K_c}{L}(\frac{\rho R_c}{\omega_z L} - 1) \qquad (48)$$

Compared with the type II compensator where a high frequency pole is included, the PI compensator does not have a high frequency pole. The boundary condition (15), expected to be accurate without the high frequency pole, becomes

$$D - \frac{1}{2} + (\frac{1 - 2D + 2D^2}{4})T(\omega_z - \frac{\rho R_c}{L}) \approx \frac{\dot{h}(d)L\omega_z}{v_s R_s K_c} \qquad (49)$$

which agrees with the boundary condition derived in [31, Eq. 4.6] (based on the describing function method where *the effect of $R_c$ is ignored* and $\rho R_c \ll \omega_z L$ is assumed) The boundary condition is rearranged as

$$D - \frac{1}{2} + (\frac{1 - 2D + 2D^2}{4})T\omega_z \approx \frac{\dot{h}(d)L\omega_z}{v_s R_s K_c} \qquad (50)$$

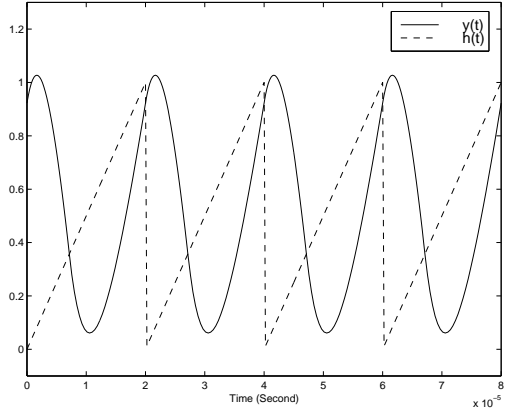

Figure 22: *Stable T-periodic* $y^0(t)$ *with a* larger *ripple*, $\omega_p = 0.81\omega_s$

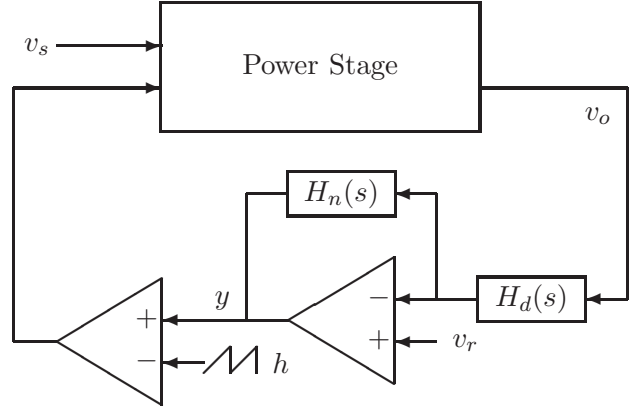

Figure 23: A DC-DC converter with a type III compensator $G_c(s) = H_n(s)/H_d(s)$

For $T(\omega_z - \rho R_c/L) \ll 1$ and from (49), the ramp slope required to avoid the subharmonic oscillation is

$$\dot{h}(d) > \frac{v_s R_s K_c}{L\omega_z}(D - \frac{1}{2}) \tag{51}$$

### 5.8 VMC with a Type III Compensator

A type III compensator [29, p. 261] has three poles, two zeros, and a gain $K_c$, with a transfer function

$$G_c(s) = \frac{H_n(s)}{H_d(s)} = \frac{K_c(1 + \frac{s}{z_1})(1 + \frac{s}{z_2})}{(s+\delta)(1 + \frac{s}{p_1})(1 + \frac{s}{p_2})} \tag{52}$$

where $\delta$ is a small number close to zero. For $\delta = 0$, the compensator has an integrator with a pole at zero.

For a buck converter with a type III compensator as shown in Fig. 23, let the state be $x =$

$(i_L, v_C, v_{e1}, v_{e2}, v_{e3})'$, where $v_{e1}$, $v_{e2}$, and $v_{e3}$ are the states of the type III compensator. Then,

$$A_1 = A_2 = \begin{bmatrix} \frac{-\rho R_c}{L} & \frac{-\rho}{L} & 0 & 0 & 0 \\ \frac{\rho}{C} & \frac{-\rho}{RC} & 0 & 0 & 0 \\ 0 & 0 & 0 & 1 & 0 \\ 0 & 0 & 0 & 0 & 1 \\ -p_1 p_2 \rho R_c & -p_1 p_2 \rho & -\delta p_1 p_2 & -\delta(p_1 + p_2) - p_1 p_2 & -\delta - p_1 - p_2 \end{bmatrix}$$

$$B_1 = \begin{bmatrix} \frac{1}{L} & 0 \\ 0 & 0 \\ 0 & 0 \\ 0 & 0 \\ 0 & p_1 p_2 \end{bmatrix}, \quad B_2 = \begin{bmatrix} 0 & 0 \\ 0 & 0 \\ 0 & 0 \\ 0 & 0 \\ 0 & p_1 p_2 \end{bmatrix}$$

$$C = K_c \begin{bmatrix} 0 & 0 & 1 & \frac{1}{z_1} + \frac{1}{z_2} & \frac{1}{z_1 z_2} \end{bmatrix}, \quad D = \begin{bmatrix} 0 & 1 \end{bmatrix}$$

$$E_1 = E_2 = \begin{bmatrix} \rho R_c & \rho & 0 & 0 & 0 \end{bmatrix}$$

Since the integrator is modeled as $1/(s + \delta)$, both $A_1$ and $A_2$ are invertible.

**Case I:** $p_1 > \omega_s/10$.

A typical guideline [29, p. 412] popular in industry to set the parameters of the compensator is as follows. Set one pole at $\delta \approx 0$ (as an integrator), and set $p_1 = \omega_s/2$ and $p_2 = 1/R_c C$. Set the gain $K_c$ to adjust the phase margin and the crossover frequency. Let $z_1 = \kappa_z/\sqrt{LC}$ and $z_2 = 1/\sqrt{LC}$, where $\kappa_z$ is a zero scale factor to have additional flexibility to adjust the phase margin and the crossover frequency. The zero scale factor $\kappa_z$ used in industry typically varies between 0.1 and 1.2. As will be shown later, a smaller value of $\kappa_z$ may lead to the subharmonic oscillation. Taking into account the above guidelines, the compensator has a transfer function

$$G_c(s) = \frac{K_c(1 + \frac{\sqrt{LC}s}{\kappa_z})(1 + \sqrt{LC}s)}{(s + \delta)(1 + \frac{2s}{\omega_s})(1 + R_c C s)} \tag{53}$$

The next example shows that, with the compensator (53), the subharmonic oscillation still occurs even with a phase margin of 38.9 degrees. As in ACMC, one compensator pole is close to $\omega_s$, and those higher order terms ($CA_1^2 B_{11} T^2$, for example) in (17) may be larger than $CB_{11}$ and $CA_1 B_{11} T$, and one needs to use the exact condition (13).

**Example 4.** (*With phase margin of 38.9 degrees, the subharmonic oscillation still occurs.*) Consider a buck converter with the type III compensator (53). Exactly the same parameters as in [32] are used: $f_s = 1/T = 300$ kHz, $L = 900$ nH, $C = 990$ μF, $R = 0.4$ Ω, $R_c = 5$ mΩ, $v_s = 5$ V, $v_r = 3.3$ V, $V_h = 1.5$ V, $K_c = 7.78 \times 10^4$, $z_1 = 1/2\sqrt{LC} = 1.675 \times 10^4$, $z_2 = 1/\sqrt{LC} = 3.35 \times 10^4$, $p_1 = \omega_s/2 = 9.425 \times 10^5$, and $p_2 = 1/R_c C = 2.02 \times 10^5$.

Simulation (Fig. 24) shows that the subharmonic oscillation occurs when $v_s = 16$ V ($D \approx 0.206$). This is also confirmed by the exact sampled-data analysis [21] with a sampled-data pole at -1 when the subharmonic oscillation occurs. The loop gain frequency response (Fig. 25) shows that even with a phase margin of 38.9 degrees, the subharmonic oscillation still occurs.

The subharmonic oscillation can be accurately predicted by the slope-based analysis. If $D$ is designed or known as 0.2, for example, then $V_s^* = 15.6$ based on (13), which is close to the simulation results and the exact sampled-data analysis. Accuracy can be improved by using (13) and (31) to determine the right duty cycle, shown in Fig. 26, which shows exactly that $V_s^* = 16$ and $D = 0.206$. □

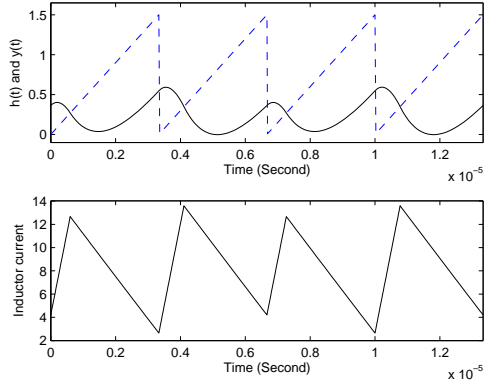

Figure 24: Signal waveforms showing the subharmonic oscillation

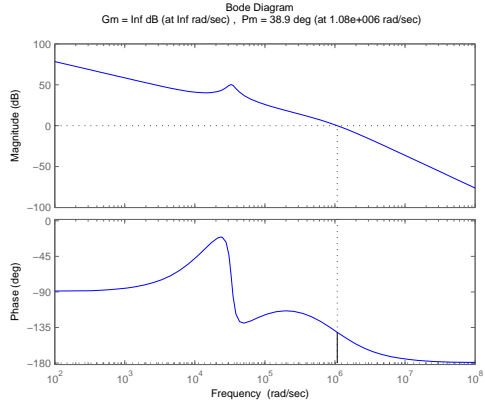

Figure 25: Loop gain frequency response showing phase margin of 38.9 degrees, but the subharmonic oscillation still occurs.

**Example 5.** (*Unstable window of $p_1$, unrelated to the ripple size of $y^0(t)$.*) Consider again Example 4. Instead of $v_s = 5$, let $v_s = 16$, the subharmonic oscillation occurs. The stable $2T$-periodic orbit $y^0(t)$ is similar to Fig. 24, omitted to save space. The *unstable $T$-periodic orbit $y^0(t)$* is shown in Fig. 27.

Here, $p_1 = 0.5\omega_s$. In the following, $p_1$ is varied from $0.1\omega_s$ to $0.6\omega_s$ to see the effect of this pole location on stability. Similar to Example 3, the value of $p_1$ adjusts the ripple size of $y^0(t)$. A larger $p_1$ leads to a larger ripple of $y^0(t)$. In [9], it is hypothesized that the ripple size of $y^0(t)$ is related to subharmonic oscillation. The following simulation shows that the ripple size of $y^0(t)$ is *unrelated* to subharmonic oscillation. The S plot (Fig. 28) shows an unstable window of $p_1 \in (0.23, 0.5)\omega_s$. For $0.23\omega_s < p_1 < 0.5\omega_s$, the S plot is above $\dot{h}(d) = 450000$ and the converter is unstable with subharmonic oscillation. The unstable window of $p_1$ is confirmed by time simulation. For $p_1 = 0.2\omega_s$, the ripple size of $y$ is small, and $y^0(t)$ is stable (Fig. 29). For $p_1 = 0.5\omega_s$, the ripple size of $y$ is larger, and $y^0(t)$ is unstable (Fig. 27). For $p_1 = 0.6\omega_s$, the ripple size of $y$ is even larger, but $y^0(t)$ is stable (Fig. 30). Comparing Figs. 29-30, the ripple size of $y^0(t)$ is unrelated to subharmonic oscillation. This shows a counter-example for the hypothesis proposed in [9] that the ripple size of $y^0(t)$ is related to subharmonic oscillation.

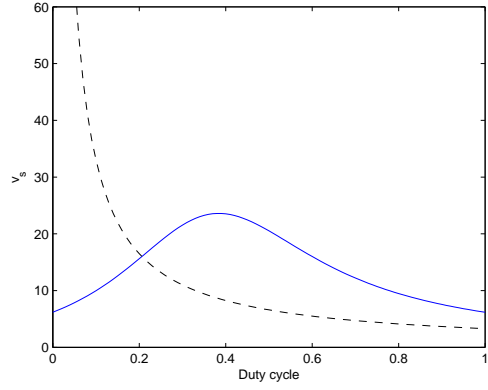

Figure 26: Curves of (13) (solid line) and (31) (dashed line), the intersection is the subharmonic oscillation condition

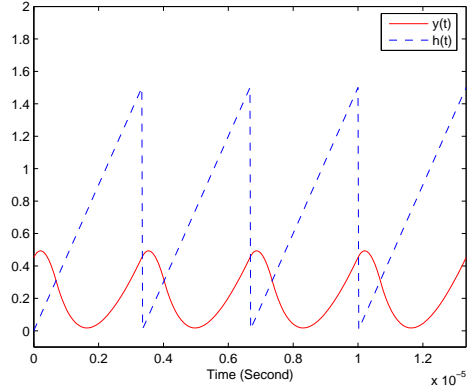

Figure 27: *Unstable $T$-periodic $y^0(t)$, $p_1 = 0.5\omega_s$.*

The unstable window of $p_1$ is also confirmed by the sampled-data pole trajectories. The sampled-data pole trajectories for $0.1\omega_s < p_1 < 0.6\omega_s$ are shown in Fig. 31. Three poles are fixed around 0.9485, 0.8853, and 0.51. A pole leaves the unit circle through -1 when $p_1 = 0.23\omega_s$, and enters the unit circle when $p_1 = 0.5\omega_s$. This explains exactly the unstable window of $p_1$. In this example, the unstable window of $p_1$ is verified by three different approaches: time simulation, the S plot, and the sampled-data pole trajectories. □

**Case II:** $p_1 < \omega_s/10$.

Similar to ACMC and based on simulations, for $\omega_p < \omega_s/10$, the boundary condition (16) is still a good approximation. Here, $CB_{11} = 0$ and $CA_1B_{11} = -p_1p_2\rho R_c K_c/z_1z_2L = -\rho p_2 K_c/\kappa_z$, then the boundary condition (16) becomes

$$V_s^* \approx \frac{4V_h \kappa_z}{T^2 \rho p_2 K_c}\left(\frac{1}{1 - 2D + 2D^2}\right) \tag{54}$$

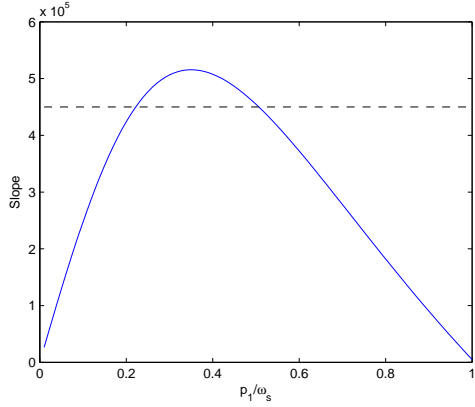

Figure 28: The intersection of the S plot (solid line) and $\dot{h}(d)$ (dashed line) shows the unstable window of $p_1 \in (0.23, 0.5)\omega_s$.

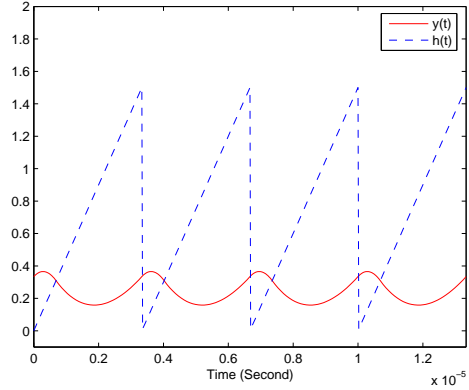

Figure 29: Stable $T$-periodic $y^0(t)$, $p_1 = 0.2\omega_s$.

# 6 Harmonic Balance Analysis of the Buck Converter

A brief review of harmonic balance (HB) analysis is presented. The HB analysis is useful for *both* steady-state analysis and small-signal analysis of nonlinear systems [10, 18, 19, 20]. Since subharmonic oscillation of the boost converter under PVMC in *continuous* conduction mode does not occur as discussed in Sec. 4.3, without loss of generality, only the buck converter is considered.

Consider a buck converter power stage, with a control-to-output ($D$-to-$v_o$) transfer function $G_{vd}(s)$. In the converter, there is an ON switch and an OFF switch (sometimes substituted by a diode). Let the voltage across the OFF switch (or the diode) be $v_d$ (as shown in Fig. 5, for example). The waveform of $v_d(t)$ is a square wave with the high voltage at $v_s$ and the low voltage at 0, which can be represented by Fourier series (harmonics).

In the converter, some parts are linear (from $v_d$ to $y$) and some are nonlinear (from $y$ to $v_d$). Let the $v_d$-to-$v_o$ transfer function be $G_v(s)$. One has [1, p. 470]

$$G_v(s) = \frac{G_{vd}(s)}{v_s} = \frac{sR_cC + 1}{LC(1 + \frac{R_c}{R})s^2 + (\frac{L}{R} + R_cC)s + 1} \tag{55}$$

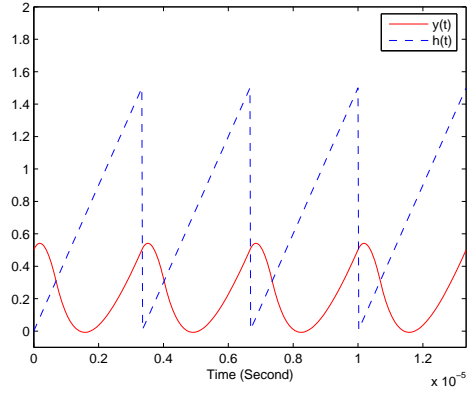

Figure 30: *Stable T*-periodic $y^0(t)$ with a *larger* ripple, $p_1 = 0.6\omega_s$.

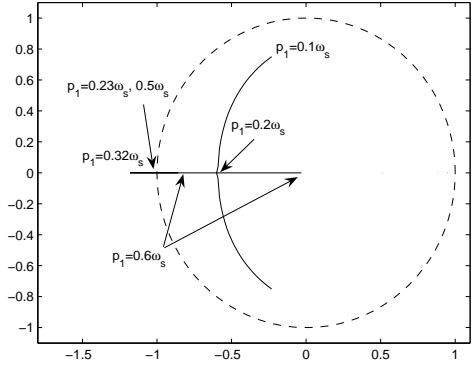

Figure 31: Sampled-data pole trajectories for $0.1\omega_s < p_1 < 0.6\omega_s$.

Let the compensator transfer function (from $v_o$ to $-y$ (negative sign due to the negative feedback)) be $G_c(s)$. Let the transfer function from $v_d$ to $-y$ be $G(s)$. Then, for VMC, $G(s) = G_c(s)G_v(s) = G_c(s)G_{vd}(s)/v_s$. Note that, for CMC, $G(s)$ has a similar form as discussed later. Assume the PWM modulator has a gain $1/V_h$. Let the loop gain be $T(s)$, then $T(s) = G_c(s)G_{vd}(s)/V_h$. The gain $G(s)$ is proportional to the loop gain by

$$G(s) = \frac{V_h}{v_s}T(s) \qquad (56)$$

The intersection of $h(t)$ with the T-periodic solution $y^0(t) = Cx^0(t) + Du$ determines the duty cycle and hence the waveform of $v_d(t)$. By "balancing" the equation $y^0(t) = h(t)$ (written in Fourier series form) at the switching instants, conditions for existence of periodic solutions and subharmonic oscillation can be derived. Let **Re** denote taking the real part of a complex number. Based on [10, 20], a necessary and sufficient condition for the occurrence of subharmonic oscillation in a buck converter (with the trailing edge modulation) is $v_s < V_s^*$, where

$$V_s^* = \frac{V_h}{2\mathbf{Re}\left[\sum_{k=1}^{\infty}[(1 - e^{j2k\pi D})G(jk\omega_s) - G(j(k - \frac{1}{2})\omega_s)]\right]} \quad (57)$$

expressed in terms of $V_h$, $D$, and the transfer function $G(s)$ evaluated at half the switching frequency and its harmonics. Since both (13) and (57) are exact conditions for the occurrence of subharmonic oscillation, it can be shown that (13) and (57) are equivalent, but expressed in different forms.

The boundary condition (57) also leads to a different expression of the S plot. From (57),

$$2v_s f_s \mathbf{Re}[\sum_{k=1}^{\infty}((1 - e^{j2\pi kD})G(j\omega_s k) - G(j\omega_s(k - \frac{1}{2})))] = \dot{h}(d) \quad (58)$$

where the left side of (58) is the S plot expressed in terms of harmonics. One can prove that (58) is equivalent to (12), where the S plot is expressed in terms of matrices.

Generally, $G_v(s)$, $G_c(s)$ and thus $G(s) = G_c(s)G_v(s)$ are low-pass filters. The denominator of (57) can be approximated by the term that involves $G(s)$ with the smallest argument. Therefore, (57) becomes

$$V_s^* \approx \frac{V_h}{2\mathbf{Re}\left[(1 - e^{j2\pi D})G(j\omega_s) - G(\frac{j\omega_s}{2})\right]} \quad (59)$$

The HB analysis is applied to the various control schemes discussed above. For switching frequency much larger than $1/\sqrt{LC}$ and $1/RC$, the condition (57) can be further simplified and expressed in closed forms. The derived closed-form boundary conditions are almost identical to those based on the sampled-data slope-based analysis discussed above, further corroborating the accuracy of the derived conditions.

## 6.1 Proportional Voltage Mode Control (PVMC)

Here, $G_c(s) = k_p$ and $G(s) = G_c(s)G_v(s) = k_p G_v(s)$. Based on the facts that for $0 < D < 1$

$$\sum_{k=1}^{\infty} \frac{1 - \cos(2\pi kD)}{k^2} = \pi^2 D(1 - D) \quad (60)$$

$$\sum_{k=1}^{\infty} \frac{2}{(k - \frac{1}{2})^2} = \pi^2 \quad (61)$$

$$\sum_{k=1}^{\infty} \frac{\sin(2\pi kD)}{k} = \pi(\frac{1}{2} - D) \quad (62)$$

then from (55) and (57),

$$V_s^* \approx \frac{V_h LC\omega_s^2}{k_p \rho}\left(\frac{1}{2\sum_{k=1}^{\infty}(-\omega_s R_c C(\frac{\sin(2\pi kD)}{k}) + \frac{1}{(k-\frac{1}{2})^2} + \frac{\cos(2\pi kD)-1}{k^2})}\right)$$

$$= \frac{4V_h LC}{k_p T^2 \rho}\left(\frac{1}{\frac{4R_c C}{T}(D - \frac{1}{2}) + 1 - 2D + 2D^2}\right) \quad (63)$$

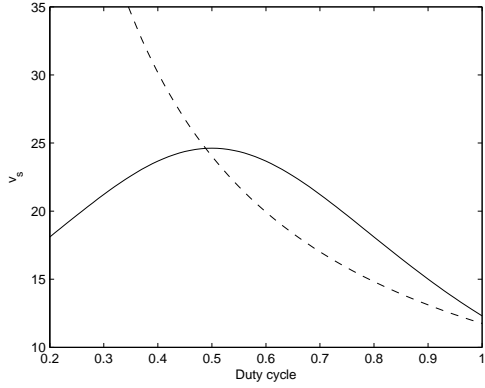

Figure 32: The intersection of the two curves (63) (solid line) and (31) (dashed line) shows $V_s^* = 24.5$

which is similar to (28) (based on the slope-based analysis) for $R_c \ll R$ (which results in $\rho \approx 1$) and $R_c^2 \ll L/C$. Since the harmonic balance analyzes the converter in the frequency domain, it has some advantages over the slope-based analysis. For example, it can show how the loading $R$ affects the subharmonic oscillation as discussed later.

**Example 6.** Consider a widely studied buck converter in [24]. The converter parameters are $T = 400$ μs, $L = 20$ mH, $C = 47$ μF, $R = 22$ Ω, $V_h = 4.4$, and $k_p = 8.4$. To transform from the leading-edge modulation to the generally used trailing-edge modulation and to remove the ramp offset in [24], let $v_r = 12.276$, which can be proved to generate equivalent results as in [24]. Subharmonic oscillation is known to occur with $V_s^* = 24.5$. The intersection of the two curves (63) and (31) also shows $V_s^* = 24.5$ in Fig. 32. □

**Dependence on loading resistance $R$.**
In (63), $V_s^*$ is almost independent of $R$ because $\rho = R/(R + R_c) \approx 1$ for $R_c \ll R$. For practical converters, $\omega_s > 1/RC$, and the frequency ratio $\tau := 1/RC\omega_s$ is small. Then, (63) is a good approximation. For large $\tau$, the dependence of $V_s^*$ on $R$ becomes significant, which is studied next. Without loss of generality, assume that $R_c/R$ is small.

From (55) and (57),

$$V_s^* \approx \frac{V_h LC\omega_s^2}{2k_p}\left(\frac{1}{2}\sum_{k=1}^{\infty}\left(\frac{1}{(k-\frac{1}{2})^2 + \tau^2} + \frac{\cos(2\pi kD) - 1 - \frac{\tau}{k}\sin(2\pi kD)}{k^2 + \tau^2}\right)\right) \quad (64)$$

$$\approx \frac{1}{\frac{1}{4} + \tau^2} + \frac{\cos(2\pi D) - 1 - \tau\sin(2\pi D)}{1 + \tau^2} + \frac{1}{\frac{9}{4} + \tau^2} + \frac{\cos(4\pi D) - 1 - \frac{\tau}{2}\sin(4\pi D)}{4 + \tau^2} \quad (65)$$

The next example illustrates the dependence on $R$.

**Example 7.** (*Dependence on $R$*) Consider again Example 6 which has $\tau = 0.062$. Now decrease the loading resistance to $R = 10$ Ω, then $\tau = 0.136$ becomes significant. Dependence on $R$ is reported in [33, Fig. 7], which shows $V_s^* = 26.8$. The approximate curve (65) and the exact curve (57) (also the same as (13)) are shown in Fig. 33. Based on (57), $V_s^* = 26.8$, agreed exactly with [33]. Based on (65), $V_s^* = 28$, also agreed closely with [33]. □

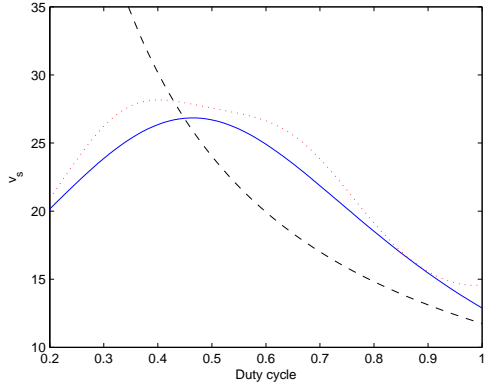

Figure 33: The intersection of the two curves (57) (solid line) and (31) (dashed line) shows $V_s^* = 26.8$. The intersection of the two curves (63) (dotted line) and (31) (dashed line) shows $V_s^* = 28$

## 6.2 CMC with the Voltage Loop Open

In CMC with the voltage loop open, the controlled output is $i_L$. Similar to (55), the $v_d$-to-$i_L$ transfer function is [1, p. 470]

$$G_i(s) := \frac{(1 + \frac{R_c}{R})Cs + \frac{1}{R}}{LC(1 + \frac{R_c}{R})s^2 + (\frac{L}{R} + R_cC)s + 1} \tag{66}$$

Since no extra compensator (except the compensating ramp $h(t)$) is added in the current loop, $G_c(s) = 1$, and one has $G(s) := G_c(s)G_i(s) = G_i(s)$ for CMC. From (57) and (62),

$$V_s^* \approx \frac{-V_h L}{T_s \rho \sum_{k=1}^{\infty} \frac{\sin(2\pi k D)}{\pi k}} = \frac{V_h L}{T_s \rho (D - \frac{1}{2})} \tag{67}$$

For $R_c \ll R$, (67) is close to (39) (based on the slope-based analysis).

## 6.3 CMC with the Voltage Loop Closed

Here, there are two feedback loops. Let $G_c(s)$ be the voltage loop compensator. In the $s$-domain, $y = i_c - i_L = G_c(s)(v_r - v_o) - i_L = G_c(0)v_r - (G_c(s)G_v(s) + G_i(s))v_d$. Then,

$$G(s) = G_c(s)G_v(s) + G_i(s) \tag{68}$$

Without loss of generality, let $G_c(s) = k_p$. Then, based on (55) and (66), the boundary condition (57) rearranged in terms of $k_p$ is

$$k_p^* \approx \frac{\frac{m_a L}{v_s} + \frac{T}{4}(\frac{1}{RC} + \frac{R_c}{L}) - D + \frac{1}{2}}{\frac{T}{4}(\frac{1-2D+2D^2}{4}) - \frac{TR_c}{4}(\frac{1}{RC} + \frac{R_c}{L}) + (D - \frac{1}{2})R_c} \tag{69}$$

which is similar to (41) (based on the slope-based analysis).

**Example 8.** Consider again Example 2. Based on (69), one has $k_p^* = 229$, agreed closely with the simulation results ($k_p^* = 237$). The small discrepancy between the HB analysis and the simulation results is due to the error of the duty cycle as discussed in Example 2. □

## 6.4 Average Current Mode Control (ACMC) with a Type II Compensator

The analysis of the ACMC buck converter in [12] is presented here for completeness. The denominator of (57) can be approximated by the term that involves $G$ with the smallest argument, then

$$V_s^* \approx \frac{V_h}{2\mathbf{Re}[G(j\omega_s) - G(\frac{j\omega_s}{2})]} \tag{70}$$

For an ACMC buck converter with $\omega_s \gg 1/\sqrt{LC}$ and $1/RC$, (70) can be simplified as

$$V_s^* \approx \frac{V_h L \omega_z f_s}{R_s K_c} \psi(\theta) \tag{71}$$

where the frequency ratio $\theta := \omega_p/\omega_s$ and

$$\psi(\theta) = \frac{\pi(1+\theta^2)(1+4\theta^2)}{3\theta} \tag{72}$$

The function $\psi(\theta) = \psi(\omega_p/\omega_s)$ has a minimum value of 5 at $\omega_p = 0.38\omega_s$. From (71), it implies that larger values of $\psi$, $V_h$, $L$, $\omega_z$, and $\omega_s$, or smaller values of $R_s$ and $K_c$ would lead to a larger stable operating range of the source voltage $v_s$. This gives insight on how these parameters affect subharmonic oscillation. A larger value of $K_c/\omega_z$ leads to larger mid-frequency gain of the current-loop compensator and larger crossover frequency of the current loop [34, 35]. However, that also leads to instability.

A design guideline proposed in [36] suggests that the value of $\omega_p$ is chosen between $0.33\omega_s$ and $0.5\omega_s$. From Fig. 20, this choice of $\omega_p$ results in smaller $V_s^*$ and hence smaller stable operating range of the source voltage. Setting $\omega_p = \omega_s$ instead, for example, will have larger operating range of the source voltage. From (71), $V_s^*$ is a function of $\omega_p$. As $\omega_p$ affects stability nonlinearly, it is desirable to remove such effect (dependence on $\omega_p$). Let

$$V_s^*|_{min} := V_s^*|_{\theta=0.38} \approx \frac{5 V_h L \omega_z f_s}{R_s K_c} \tag{73}$$

which is not a function of $\omega_p$. If $v_s < V_s^*|_{min}$, the converter is free from period-doubling for all $\omega_p$. If $v_s \geq V_s^*|_{min}$, then there exists an unstable (period-doubling) window of $\omega_p$, and the converter is stable by choosing a larger value of $\omega_p$. Note that $V_s^*|_{min}$ is the *minimum* critical source voltage for *all* $\omega_p$. It is possible that for a particular $\omega_p$ and $v_s \geq V_s^*|_{min}$, the converter is still stable. But to eliminate the dependence on $\omega_p$, $V_s^*|_{min}$ is the *maximum* allowable source voltage with stability for all $\omega_p$.

Another *stability* guideline proposed in [28] suggests, expressed in terms of $v_s$,

$$v_s \leq \min[\frac{2}{1-D}, \frac{1}{D}]\frac{V_h L \omega_z f_s}{R_s K_c} < \frac{3 V_h L \omega_z f_s}{R_s K_c} \tag{74}$$

This stability guideline is more conservative than (73). The converter may be stretched as in (73) to achieve better dynamics.

## 6.5 VMC with a Type III Compensator

With the compensator (53), which cancels some poles and zeros of the power stage, one has

$$G(s) = G_c(s) G_v(s) \approx \frac{K_c}{\kappa_z s(1 + \frac{2s}{\omega_s})} \tag{75}$$

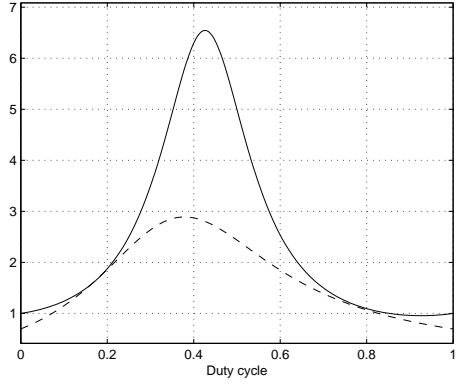

Figure 34: Plot of $\phi(D)$, solid line for (78) and dashed line for more accurate (77)

which depends on $\omega_s$, and the compensator parameters $\kappa_z$ and $K_c$. From (57) and (75), the subharmonic oscillation is avoided if and only if

$$v_s < V_s^* = \frac{V_h \omega_s \kappa_z}{2K_c} \phi(D) \tag{76}$$

where $\phi(D)$ is

$$\phi(D) = \frac{1}{\mathbf{Re}\left[\sum_{k=1}^{\infty} \frac{1-e^{j2k\pi D}}{k(j-2k)} - \frac{1}{(k-\frac{1}{2})(j-2k+1)}\right]} \tag{77}$$

A smaller $\kappa_z$ or a larger $K_c$ would result in a smaller $V_s^*$. Generally, $\omega_s$ is large, and it may seem that the condition (76) is generally met and the subharmonic oscillation is avoided. However, a large value of $K_c$ (in the order of $\omega_s$) may result in $v_s > V_s^*$ and therefore subharmonic oscillation occurs. If (59), instead of (57), is used, the following approximate expression for $\phi(D)$ is obtained

$$\phi(D) \approx \frac{5}{3 + 2\cos(2\pi D) - \sin(2\pi D)} \tag{78}$$

Both (77) and (78) are shown in Fig. 34.

**Example 9.** (*Effects of zero location*) Consider again Example 4 where $z_2 = 1/2\sqrt{LC}$ (or $\kappa_z = 1/2$), and one has $V_s^* = 16$. Now change $z_2$ to $1/\sqrt{LC}$ (or $\kappa_z = 1$) to see its effects. Simulation, confirmed with the exact sampled-data analysis, shows that the subharmonic oscillation occurs at $V_s^* = 23.9$ V (and $D \approx 0.138$). Compared with Example 4, one sees that a larger $\kappa_z$ results in wider operating range of the source voltage. This effect can be clearly seen in (76) which shows that a larger $\kappa_z$ results in a larger $V_s^*$. The value of $V_s^*$ is not linearly proportional to $\kappa_z$ because a different $v_s$ results in a different duty cycle to regulate to the same output voltage, and thus a different value of $\phi(D)$.

The subharmonic oscillation can be accurately predicted by the harmonic balance analysis (or equivalently the slope-based analysis). The plots of (76) (same as (13)) and (31) are shown in Fig. 35, intersecting exactly at $(D, V_s^*) = (0.138, 23.9)$. Compared with Fig. 26, the curve (76), but not $V_s^*$, is multiplied by two because $\kappa_z$ is also multiplied by two. □

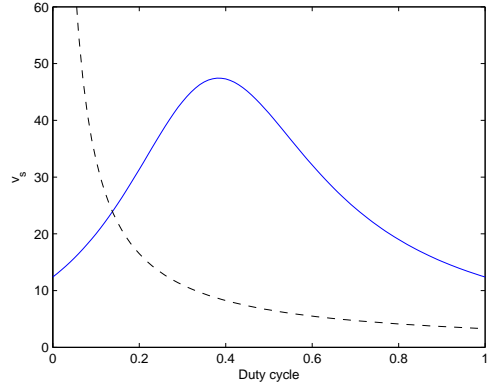

Figure 35: Curves of (76) (solid line) and (31) (dashed line), the intersection is the subharmonic oscillation condition

## 7 Prediction of Subharmonic Oscillation Based on Loop Gain

Since the gain $G(s)$ is proportional to the loop gain $T(s)$ by (56), and the subharmonic oscillation is avoided if and only if $v_s < V_s^*$, where $V_s^*$ is given in (57), which directly (by simple algebra) leads to the following theorem. This theorem shows that a limitation on the loop gain is required to avoid the subharmonic oscillation.

**Theorem 1** *Given a closed-loop buck converter with a loop gain $T(s)$, subharmonic oscillation is avoided if and only if*

$$\mathbf{Re}\left[\sum_{k=1}^{\infty}(1 - e^{j2k\pi D})T(jk\omega_s) - T(j(k - \frac{1}{2})\omega_s)\right] < \frac{1}{2} \tag{79}$$

The traditional loop gain analysis to design a converter with enough gain and phase margins only helps to avoid the Neimark-Sacker and the saddle-node bifurcations, but not the subharmonic oscillation, unless the system dimension is increased by considering the sampling effect as in [4, 5, 6]. This theorem, putting an additional limitation on the familiar loop gain, helps to fill this gap.

Using only one term in the series summation, (79) becomes

$$\mathbf{Re}\left[(1 - e^{j2\pi D})T(j\omega_s) - T(j\frac{\omega_s}{2})\right] < \frac{1}{2} \tag{80}$$

As the following analysis indicates, a conservative design to make the converter free from the subharmonic oscillation *for all duty cycles* is by setting $D = 1$ in (80). Then (80) becomes

$$\mathbf{Re}\left[T(j\frac{\omega_s}{2})\right] > -\frac{1}{2} \tag{81}$$

which can be expressed in terms of Nyquist plot. In the Nyquist plot, the real part of the loop gain at half the switching frequency needs to be greater than $-1/2$ to avoid the subharmonic oscillation for all duty cycles. Therefore, the Nyquist plot is not only helpful to predict the Neimark-Sacker and the saddle-node bifurcations, it is also helpful to predict the subharmonic oscillation based on (81). Since the Nyquist plot is widely used, (81) is helpful to design a stable converter. However, it should be noted that (81) is only an approximate condition, and the exact condition is (79).

**"HB plot": a Nyquist-like plot in the complex plane.**
Note that the exact condition (79) is a function of $D$, $\omega_s$, and the loop gain $T(s)$, where $T(s)$ is further a function of $v_s$, $V_h$, and the power stage and compensator parameters. Without loss of generality, the following design procedure for the line regulation design is proposed. In the line regulation design, one wants to find the stable operating range of $v_s$. Given an desired output voltage, this is equivalent to find the stable operating range of $D$. Similar design procedures can be applied to determine the operating range of *other parameters*. Let

$$H(D) := \sum_{k=1}^{\infty}(1 - e^{j2k\pi D})T(jk\omega_s) - T(j(k-\frac{1}{2})\omega_s) \tag{82}$$

Then, the boundary condition (79) to avoid the subharmonic oscillation becomes

$$\mathbf{Re}\left[H(D)\right] < \frac{1}{2} \tag{83}$$

For designation purpose, $H(D)$ is called an HB plot because it is similar to the Nyquist plot to facilitate the converter design. Generally the feedback gain is large for a *practical* buck converter, then $v_s \approx v_r/D$. This value of $v_s$ may be used for the HB plot. Given a desired range of $D$, one can plot $H(D)$ according to (82). Those values $D$ such that the HB plot is on the left of $1/2$ in the complex plane are the stable operating range of $D$ to avoid the subharmonic oscillation.

**"M plot": a similar HB plot in matrix representation.**
Since both (13) and (57) are exact conditions for occurrence of subharmonic oscillation, they are actually identical. From (13) and (57),

$$\mathbf{Re}\left[\sum_{k=1}^{\infty}[(1 - e^{j2k\pi D})G(jk\omega_s) - G(j(k-\frac{1}{2})\omega_s)]\right] = \frac{T}{2}C[(I - e^{A_1T})^{-1}(e^{A_1d} - I) + (I + e^{A_1T})^{-1}]B_{11} \tag{84}$$

Based on (13) and (28), and similar to the HB plot in the *complex plane*, define a new "M plot" in the matrix form and as a function of $D$ in the *real domain*,

$$M(D) := \frac{Tv_s}{V_h}C[(I - e^{A_1T})^{-1}(e^{A_1TD} - I) + (I + e^{A_1T})^{-1}]B_{11} \tag{85}$$

$$\approx \frac{k_pT^2v_s}{4V_hLC}[\frac{4R_cC}{T}(D-\frac{1}{2}) + \rho(1-\frac{R_c^2C}{L})(1-2D+2D^2)] \tag{86}$$

where $v_s \approx v_r/D$ for a large feedback gain (which is true for most practical converters). Note that (85) is an exact representation, and (86) is an approximate representation, which is generally accurate if the controller poles are smaller than $\omega_s/10$ as discussed previously. Then, from *Theorem 1*, (56) and (84), one has

**Corollary 1** Given a closed-loop buck converter as in Fig. 1, subharmonic oscillation is avoided if and only if

$$M(D) < 1 \tag{87}$$

Actually, this corollary can be also proved easily from (13) (or (28)), by dividing both sides of (13) (or (28)) by the right side (to normalize the right side to one). Both (85) and (13) are equivalent exact boundary conditions but expressed in different forms.

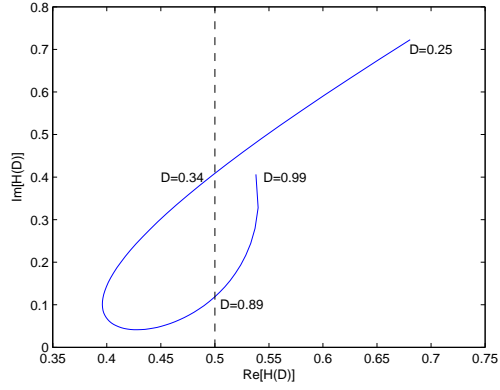

Figure 36: HB plot in the complex plane showing the stable operating range of $D = [0.34, 0.89]$.

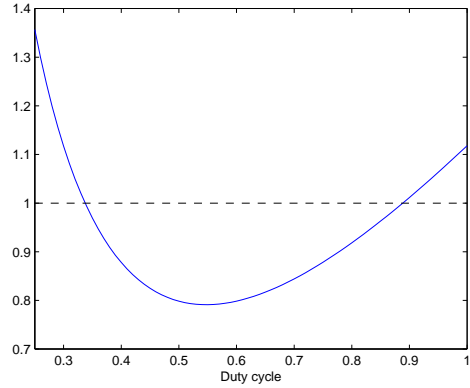

Figure 37: M plot $M(D)$ showing the stable operating range of $D = [0.34, 0.89]$.

**Example 10.** (*Line regulation: determine the stable operating ranges of $D$ and $v_s$, based on the HB plot and the M plot.*) Consider again Example 1, but with $R_c = 2$ mΩ. In Example 1, it is shown that the stable operating range of $D$ is $[0.34, 0.89]$. Here, using the HB plot and the M plot will confirm the same result.

An HB plot for $D = [0.25, 0.99]$ is shown in Fig. 36, intersecting with the vertical line at $1/2$ in the complex plane at $D = 0.34$ and $0.89$. An M plot for $D = [0.25, 0.99]$ is shown in Fig. 37, intersecting with the horizontal line at 1 at $D = 0.34$ and $0.89$. Both the plots agrees exactly with the analysis in Example 1. Given an output voltage around $v_r = 4$, the corresponding stable operating range of $v_s$ free from the subharmonic oscillation is $[4/0.89, 4/0.34] = [4.48, 11.85]$. □

**Example 11.** (*Application of Theorem 1.*) Consider again Example 6, but with $R = 2$ Ω and $v_s = 50$ V. Using the *exact* sampled-data analysis, $D = 0.243$, $x^0(0) = (5.9867, 12.0753)'$, and $x^0(d) = (6.1711, 12.1486)'$. The eigenvalues of $\Phi$ are -0.0336 and -0.4222, and they are stable. The steady-state simulation is shown in Fig. 38, also indicating that the converter is stable. From (82), $H(0.243) = 0.1390 + 0.8867j$ and $\mathbf{Re}\,[H(0.243)] < 1/2$. Based on Theorem 1, the converter is stable, agreed with the sampled-data analysis and the simulation. From (13), subharmonic

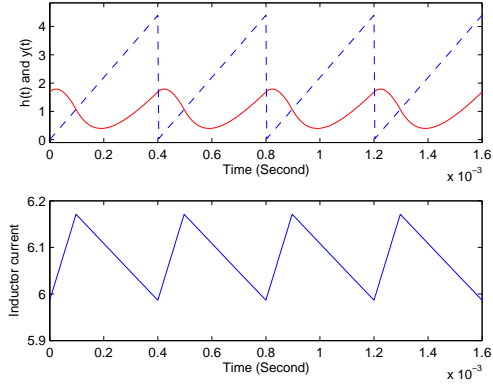

Figure 38: Plots of $y(t)$ and $i_L(t)$ (solid lines), and $h(t)$ (dashed line), $R = 2$ and $v_s = 50$

oscillation occurs when $v_s > 82.9$. □

# 8 Subharmonic Oscillation Caused by a Large Crossover Frequency

A rule of thumb to select the crossover frequency $\omega_c$ is by setting $\omega_c < \omega_s/5$ [29, p. 250]. No theoretical explanation is given. The following analysis will give a theoretical explanation and puts a limit on the crossover frequency to avoid the subharmonic oscillation. Since the compensator poles and zeros are selected based on the power stage parameters, such as $L$, $C$, and $R_c$, the selected poles and zeros may result in a large crossover frequency, and the subharmonic oscillation occurs.

The loop gain analysis in *Theorem 1* can lead to a condition based on the crossover frequency. Without loss of generality, only two control schemes are considered. Similar analysis can be applied to other control schemes.

## 8.1 Buck Converter with a Type III Compensator

**Case I:** $p_1 = \omega_s/2 > \omega_s/10$.
First, a closed-form crossover frequency is derived. The loop gain (after some poles and zeros being canceled) for the frequency below $\omega_s/2$ is approximately

$$T(s) = G(s)\frac{v_s}{V_h} \approx \frac{K_c v_s}{\kappa_z s(1 + \frac{2s}{\omega_s})V_h} \approx \frac{K_c v_s}{s\kappa_z V_h} \tag{88}$$

From (88), the crossover frequency $\omega_c$ is around $K_c v_s/\kappa_z V_h$. In Example 9, it is shown that a smaller $\kappa_z$ leads to a smaller operating range of $v_s$. From (88), a smaller $\kappa_z$ leads to a larger $\omega_c$, and the subharmonic oscillation is more likely to occur as discussed next. With the estimated crossover frequency $\omega_c^{est} := K_c v_s/\kappa_z V_h$, rearranging (76) leads to the following theorem.

**Theorem 2** *In a buck converter with a type III compensator (53), the subharmonic oscillation is avoided if and only if*

$$\omega_c^{est} := \frac{K_c v_s}{\kappa_z V_h} < \frac{\omega_s \phi(D)}{2} \tag{89}$$

*where $\phi(D)$, shown in (77), varies from 0.694 to 2.89 (Fig. 34).*

A larger $K_c v_s$ or a smaller $\kappa_z V_h$ leads to a larger $\omega_c$ to a point such that (89) is violated and the subharmonic oscillation occurs. Since the minimum of $\phi(D)$ is 0.694, a rule of thumb to avoid the subharmonic oscillation *for all duty cycles* is $\omega_c < 0.347\omega_s$. This theorem gives a theoretical explanation on the traditional wisdom on setting $\omega_c < \omega_s/5$. In Example 4, with $\kappa_z = 0.5$, the subharmonic oscillation occurs when $\omega_c = 1.08 \times 10^6 = 0.58\omega_s$ (shown in Fig. 25), where $\omega_c$ is too large.

From Fig. 34, the maximum of $\phi(D)$ occurs around $D = 0.4$ with a value of 2.89. Therefore, around $D = 0.4$, only $\omega_c < 1.44\omega_s$ is required. The limitation on the crossover frequency for operating at this duty cycle may be relaxed. However, to have a wider operating range of $D$, $\omega_c < 0.347\omega_s$ is probably a better design (although it may be conservative in some situations).

**Case II:** $p_1 < \omega_s/10$.

As the analysis above, the approximate crossover frequency is derived first. The loop gain for frequency below $\omega_s/2$ is approximately

$$T(s) = G(s)\frac{v_s}{V_h} \approx \frac{K_c v_s}{\kappa_z s(1 + \frac{s}{p_2})V_h} \tag{90}$$

Solving $|T(jw_c)| = 1$ gives

$$w_c \approx -\frac{p_2}{2} + \sqrt{\frac{p_2^2}{4} + \frac{p_2 v_s K_c}{\kappa_z V_h}} \tag{91}$$

For $p_2 < \omega_s/10$, the approximate condition (54) is close to the exact condition (13) as discussed above. The curve of the condition (54) has a minimum at $D = 1$ (or $D = 0$) with a value of

$$V_s^*|_{min} = V_s^*|_{D=1} = \frac{4V_h \kappa_z}{T^2 p_2 K_c} \tag{92}$$

A conservative design to avoid subharmonic oscillation for *all duty cycles* is $v_s < V_s^*|_{min}$. Then, (91) leads to

$$w_c < -\frac{p_1}{2} + \sqrt{\frac{p_2^2}{4} + \frac{\omega_s^2}{\pi^2}} \tag{93}$$

For $p_2 = \omega_s/10$, (93) leads to $\omega_c < 0.27\omega_s$, which is required to avoid the subharmonic oscillation for all duty cycles. For $p_2 \ll \omega_s$, based on (93), $\omega_c/\omega_s < 1/\pi$ is required to avoid the subharmonic oscillation for all duty cycles.

## 8.2 ACMC with a Type II Compensator

First, the approximate crossover frequency is derived. Here, the loop gain at high frequency is

$$T(s) = \frac{v_s R_s}{V_h} G_c(s) G_i(s) \approx \frac{v_s R_s}{V_h} \frac{\frac{K_c s}{\omega_z}}{s(1 + \frac{s}{\omega_p})} \frac{Cs}{LCs^2} = \frac{v_s R_s K_c}{V_h L \omega_z s(1 + \frac{s}{\omega_p})} \tag{94}$$

Solving $|T(jw_c)| = 1$ gives

$$w_c \approx -\frac{\omega_p}{2} + \sqrt{\frac{\omega_p^2}{4} + \frac{\omega_p v_s K_c R_s}{\omega_z V_h L}} \tag{95}$$

Without loss of generality, only the case for $\omega_p < \omega_s/10$ is considered. For $\omega_p < \omega_s/10$, the approximate condition (46) is close to the exact condition (13) as shown in Fig. 21. The curve of the condition (46) has a minimum at $D = 1$ (or $D = 0$) with a value of

$$V_s^*|_{min} = V_s^*|_{D=1} = \frac{4V_h\omega_z L}{T^2 K_c R_s \omega_p} \tag{96}$$

A conservative design to avoid subharmonic oscillation for all duty cycles is $v_s < V_s^*|_{min}$. Then, (95) leads to

$$w_c < -\frac{\omega_p}{2} + \sqrt{\frac{\omega_p^2}{4} + \frac{\omega_s^2}{\pi^2}} \tag{97}$$

For $\omega_p = \omega_s/10$, based on (97), $\omega_c < 0.27\omega_s$ is required to avoid subharmonic oscillation for all duty cycles. For $\omega_p \ll \omega_s$, based on (93), $\omega_c/\omega_s < 1/\pi$ is required to avoid the subharmonic oscillation for all duty cycles. The analysis is almost identical as for the type III compensator discussed above.

## 9 Conclusion and Contributions

Based on a subharmonic oscillation boundary condition published in a PhD thesis (but not elsewhere) more than a decade ago, extended design-oriented boundary conditions are derived for general switching DC-DC converters expressed by a unified VMC/CMC block diagram model. Under the unified framework, the boundary conditions for both VMC and CMC have similar forms. One does not need to analyze VMC and CMC separately. The boundary conditions are expressed in terms of signal slopes. The well known slope-based instability criterion for CMC becomes a special case in this unified modeling approach. The derived sloped-based boundary conditions are *exact*, and can be further simplified in various approximate closed forms to facilitate the converter design. The sloped-based boundary conditions are applied to analyze various VMC/CMC control schemes of varying complexity. Among the various control schemes, those popular type II or type III controllers (with integrators included) are also analyzed.

Harmonic balance analysis is also applied to derive the *exact* subharmonic oscillation boundary conditions, which are equivalent to the sloped-based boundary conditions. It is also applied to analyze five different VMC/CMC control schemes. Based on the harmonic balance analysis, a new "HB plot" in the complex plane, similar to the Nyquist plot, is proposed to accurately predict the occurrence of the subharmonic oscillation. Another equivalent "M plot" in the real domain is also proposed, which also accurately predicts the occurrence of the subharmonic oscillation. Based on the harmonic balance analysis, the relation between the crossover frequency and the subharmonic oscillation can be clearly seen.

Both the sloped-based analysis and the harmonic balance analysis complement each other. The slope-based analysis analyzes the converter in the time domain and expresses the boundary conditions in matrix forms. The harmonic balance analysis expresses the boundary conditions in terms of (sub)harmonics of the switching frequency. Since most power stage and compensator transfer functions are low-pass filters, the harmonic balance analysis is particularly useful to analyze the converter with the controller poles close to the switching frequency.

Both the slope-based analysis and the harmonic balance analysis complement the traditional average analysis. The average analysis can predict the Niemark-Sacker bifurcation and the saddle-node bifurcation, but generally not the subharmonic oscillation, unless the sampling effects are considered. In Example 2, the converter has a phase margin of 129 degrees based on an average model, but the subharmonic oscillations still occur. The subharmonic oscillation in all examples presented can be accurately predicted by the slope-based analysis or the harmonic balance analysis.

The contributions of this paper are as follows. First, the following sixteen boundary conditions are reported here for the first time to the author's knowledge. Some of the conditions are equivalent to each other. These derived boundary conditions greatly facilitate future converter *design* to avoid subharmonic oscillation. Since there are many equations reported in this paper, the key equations are identified here for easy reference. The most general boundary condition for the subharmonic oscillation is (9). It is an *exact* condition expressed in terms of signal slopes. For the boost converter, (9) becomes (18), which is also an exact condition. For the buck converter, (9) becomes (12), which is equivalent to various forms: (13) in terms of $v_s$, (22) in terms of the S plot, (79) in terms of the loop gain, (83) in terms of the HB plot, and (87) in terms of the M plot. These boundary conditions can be further simplified to various approximate forms: (15) in terms of signal slopes, (16) in terms of $v_s$, and (80) in terms of the loop gain. For those control schemes presented, the boundary conditions can be further simplified to more compact forms, such as (28) for PVMC and $V^2$ control (with ESR considered), (37) for CMC with the voltage loop open (with ESR considered), (41) for CMC with the voltage loop closed, (46) for ACMC with a type II controller and small controller poles, and (54) for VMC with a type III controller and small controller poles.

Second, all the boundary conditions are in closed forms, one can see the effects of various converter parameters on the stability. In some control schemes, the effects of ESR, loading resistance, feedback gain, and pole/zero locations are further analyzed. For example, in CMC, ESR extends the stability beyond $D = 1/2$ as shown in (38). Also, for CMC or VMC, ESR causes the boundary conditions to have both terms $D - 1/2$ and $1 - 2D + 2D^2$.

Third, although only some control schemes are analyzed, this paper also provides a methodology based on the *general* boundary condition to analyze other converters with different control schemes to predict the subharmonic oscillation. As discussed in the Introduction, for a converter under a particular control scheme, once the converter is expressed in terms of the unified VMC/CMC block diagram model shown in Fig. 1, the boundary condition for that particular control scheme can be readily obtained.

Fourth, under this general methodology, those *recently* reported boundary conditions for PI control [9] and $V^2$ control [11] become *special cases* in view of the general boundary condition. In consideration of recent development to derive more boundary conditions for other converters or control schemes, the boundary conditions derived here greatly facilitate future research.

Fifth, new Nyquist-like design-oriented plots (such as the S, HB, M plots) to predict the occurrence of subharmonic oscillation are proposed. The S plot also shows the required ramp slope to prevent the occurrence of subharmonic oscillation.

Sixth, closed-form conditions (such as (89), (93) and (97)) relating the crossover frequency to the subharmonic oscillation are derived.